\documentclass{aa}  

\usepackage{graphicx}
\usepackage{txfonts}
\begin{document} 

\title{Deep secrets of intermediate-mass giants and supergiants\thanks{Tables 5, 6, and A1 are only available in electronic form at the CDS via anonymous ftp to cdsarc.u-strasbg.fr (130.79.128.5) or via http://cdsweb.u-strasbg.fr/cgi-bin/qcat?J/A+A/}}
\subtitle{Models with rotation seem to overestimate mixing effects on the surface abundances of C, N, and Na}

\author{R. Smiljanic\inst{1}
\and
          P. Donati\inst{2}
          \and
          A. Bragaglia\inst{2}
          \and
          B. Lemasle\inst{3}
          \and
          D. Romano\inst{2}
          }

   \institute{Nicolaus Copernicus Astronomical Center, Polish Academy of Sciences, Bartycka 18, 00-716, Warsaw, Poland \\
                 \email{rsmiljanic@camk.edu.pl}   
                 \and
   INAF - Osservatorio di Astrofisica e Scienza dello Spazio di Bologna, via Gobetti 93/3, 40129, Bologna, Italy
                  \and
                 Zentrum f\"ur Astronomie der Universit\"at Heidelberg, Astronomisches Recheninstitut, M\"onchhofstr. 12-14, 69120, Heidelberg, Germany
                 }

   \date{Received 2018; accepted }

\titlerunning{Na and Al abundances in intermediate-mass giants and supergiants}
\authorrunning{Smiljanic et al.}

% \abstract{}{}{}{}{} 
% 5 {} token are mandatory
 
  \abstract
   {Recent observational results have demonstrated an increase in the surface Na abundance that correlates with stellar mass for red giants between 2 and 3 M$_{\odot}$. This trend supports evolutionary mixing processes as the explanation for Na overabundances seen in some red giants. In this same mass range, the surface Al abundance was shown to be constant.}
   {Our main aim was to extend the investigation of the Na and Al surface abundances to giants more massive than 3 M$_{\odot}$. We sought to establish accurately whether the Na abundances keep increasing with stellar mass or a plateau is reached. In addition, we investigated whether mixing can affect the surface abundance of Al in giants more massive than 3 M$_{\odot}$.}
   {We obtained new high-resolution spectra of 20 giants in the field of 10 open clusters; 17 of these stars were found to be members of 9 clusters. The giants have masses between 2.5 M$_{\odot}$ and 5.6 M$_{\odot}$. A model atmosphere analysis was performed and abundances of up to 22 elements were derived using equivalent widths. Additionally, abundances of C, N, and O were determined using spectrum synthesis. The abundances of Na and Al were corrected for non-local thermodynamic equilibrium effects (non-LTE). Moreover, to extend the mass range of our sample, we collected from the literature high-quality C, N, O, and Na abundances of 32 Galactic Cepheids with accurate masses in the range between 3 M$_{\odot}$ and 14 M$_{\odot}$.
   }
   {The surface abundances of C, N, O, Na, and Al were compared to predictions of stellar evolution models with and without the inclusion of rotation-induced mixing. The surface abundances of most giants and Cepheids of the sample can be explained by models without rotation. For giants above $\sim$ 2.5 M$_{\odot}$, the Na abundances reach a plateau level of about [Na/Fe] $\sim$ 0.20-0.25 dex (in non-LTE). This is true for both Cepheids and giants in open clusters. Regarding Al, the non-LTE [Al/Fe] ratios are mostly close to solar and suggest that Al is not affected by the first dredge-up up to $\sim$ 5.0 M$_{\odot}$. Our results support previous works that found models with rotation to overestimate the mixing effects in intermediate-mass stars.}
   {}


   \maketitle
%
%-------------------------------------------------------------------

\section{Introduction}

Recently, \citet{2016A&A...589A.115S} have shown that, for giants with masses between 2 M$_{\odot}$ and 3-3.5 M$_{\odot}$, there is a trend of increasing surface Na abundance with increasing stellar mass. For stars with mass below 2 M$_{\odot}$, the Na abundances were not enhanced. The observed trend was attributed to internal mixing processes changing the stellar surface Na abundance during the first dredge-up. This observation solved the puzzle of surface Na overabundances in giants of open clusters \citep[e.g.][]{2005A&A...441..131C,2007AJ....134.1216J,2009ApJ...701..837S,2010A&A...511A..56P,2015MNRAS.446.3556M}.

Regarding surface Al abundances, \citet{2016A&A...589A.115S} found that giants below 3 M$_{\odot}$ had [Al/Fe] $\sim$ +0.06. Preliminary computations of deviations from the local thermodynamic equilibrium (henceforth, non-LTE effects) indicated an abundance correction of the order of $-$0.05 dex. Therefore, as the main-sequence progenitors of these giants likely had [Al/Fe] $\sim$ 0.00, it was concluded that there was no change in the surface abundance of Al during the first dredge-up. 

The behaviour of the surface Na and Al abundances in giants with masses above 3 M$_{\odot}$, however, was not well established. In the sample of \citet{2016A&A...589A.115S}, only one cluster probed this mass regime, i.e. \object{NGC 6705} \citep{2014A&A...569A..17C}. The giants in NGC 6705 were found to have a mean [Na/Fe] = +0.38, in non-LTE. This value can only be explained by models including rotation-induced mixing. At the same time, an Al overabundance of [Al/Fe] = +0.30 (in LTE) was found. Internal mixing processes could not explain such an overabundance of Al, not even with the inclusion of rotation-induced mixing, which does
not affect the surface Al abundance in giants with this mass \citep[see e.g.][]{2004MmSAI..75..347W,2012A&A...543A.108L}. The hypothesis was made that the Al enhancement was inherited from a peculiar primordial enrichment of the cloud that gave birth to the open cluster. This idea followed the arguments by \citet{2014A&A...563A..44M,2015A&A...580A..85M} that aimed to explain the peculiar enhancement in $\alpha$-elements of stars in NGC 6705. 

Thus, there was still the need to extend the analysis of surface Na and Al abundances to giants above 3 M$_{\odot}$. This extension would be important to clarify first whether the surface Na abundances keep increasing with increasing stellar mass or stay at a constant value and second whether the surface Al abundances are enhanced in this mass range. 

The study presented in \citet{2016A&A...589A.115S} was performed within the context of the \emph{Gaia}-ESO Survey \citep{2012Msngr.147...25G,2013Msngr.154...47R}. The targets observed in the \emph{Gaia}-ESO Survey include stars in about 60-70 open clusters, covering from young OB associations to very old open clusters \citep[see e.g.][]{2017arXiv171107699R}. However, in \emph{Gaia}-ESO, red clump giants are observed mainly in old open clusters with a few exceptions. Therefore, to extend the work of \citet{2016A&A...589A.115S}, we collected new spectroscopic data of giants in ten open clusters with ages below 500 Myr (where the giants have $>$2.5 M$_{\odot}$). To gain a deeper insight into the evolutionary mixing history of these giants, we discuss not only the abundances of Na and Al, but also those of C, N, and O. In addition, chemical abundances of up to other 22 elements were determined. 

Chemical abundances tracing evolutionary mixing processes in such intermediate-mass stars have been discussed in a number of works \citep[e.g.][]{1985ApJ...298..782L,1995AJ....110.2425L,2011MNRAS.417..649Z}. For example, \citet{2006A&A...449..655S} determined C, N, and O abundances in 19 luminous intermediate-mass giants. The [N/C] of these giants indicated different internal evolutionary histories. While the abundances in some giants were in agreement with standard evolutionary models\footnote{Standard models are those in which convection is the only transport mechanism driving mixing processes.}, in other giants rotation-induced mixing was needed to explain the observations. A similar conclusion was obtained in the non-LTE analysis of C and N abundances in more than 30 A- and F-type supergiants presented in \citet{2011MNRAS.410.1774L,2015MNRAS.446.3447L}. Further, oxygen isotopic rations ($^{16}$O/$^{17}$O and $^{16}$O/$^{18}$O) have been investigated in intermediate-mass giants by \citet{2015A&A...578A..33L}. The values seemed to be consistent with post-first dredge-up models. 
Surface Na overabundances in luminous supergiants were previously found to be large ([Na/Fe]\footnote{[A/B] = $\log$ [N(A)/N(B)]$_{\rm \star}$ $-$ $\log$ [N(A)/N(B)]$_{\rm\odot}$, where the abundances are given by number on a scale where the number of hydrogen atoms is 10$^{12}$.} above +0.3 dex reaching up to +0.7 dex) and suggested to be correlated with stellar mass \citep[e.g.][]{1983IzKry..66..130B,1986PASP...98..561S,1988Ap.....28..202B}. Later, \citet{1994PASJ...46..395T} showed that improved non-LTE corrections decreased the observed Na abundances to moderate values ([Na/Fe] $\sim$ +0.20). These values were shown to agree well with standard models computed by \citet{1995ApJ...451..298E} for supergiants above 5 M$_{\odot}$ and the correlation with stellar mass was no longer apparent.

Later on, using more refined analysis methods, \citet{2002A&A...389..519A} reduced the error bars associated with the Na abundances in supergiants and again identified the correlation with stellar mass. \citet{2005ApJ...622.1058D} then showed that to reproduce the behaviour of these new Na abundances in stars with $\log~g~<$ 1.5, the inclusion of rotation-induced mixing was needed.

Chemical analyses of Cepheids, on the other hand, by \citet{2005PASP..117.1173K}, \citet{2013MNRAS.432..769T}, and \citet{2014A&A...566A..37G} have mostly found mild overabundances of Na (i.e. [Na/Fe] $\sim$ +0.20) and no dependence with the pulsation period (which correlates with stellar mass). These results suggest that, while Na surface abundances are modified by the first dredge-up, there is no need to include effects of rotation-induced mixing.

Thus, in light of the contradictory literature results, readdressing the surface Na abundance in intermediate-mass stars seems needed. We focus on giants belonging to open clusters whose masses can be better constrained than in field giants. Moreover, we use literature data for Cepheids to extend the sample to higher masses. This work is organised as follows. In Sect.\ \ref{sec:data} we describe our new observations, the open clusters selected for analysis, the membership status of the observed stars, and the Cepheid's data collected from the literature. In Sect.\ \ref{sec:analysis} we present the details of the spectroscopic analysis. Afterwards, Sect.\ \ref{sec:disc} presents a discussion of our results and Sect.\ \ref{sec:end} summarises our findings.

\begin{figure*}
\centering
\includegraphics[height = 13cm]{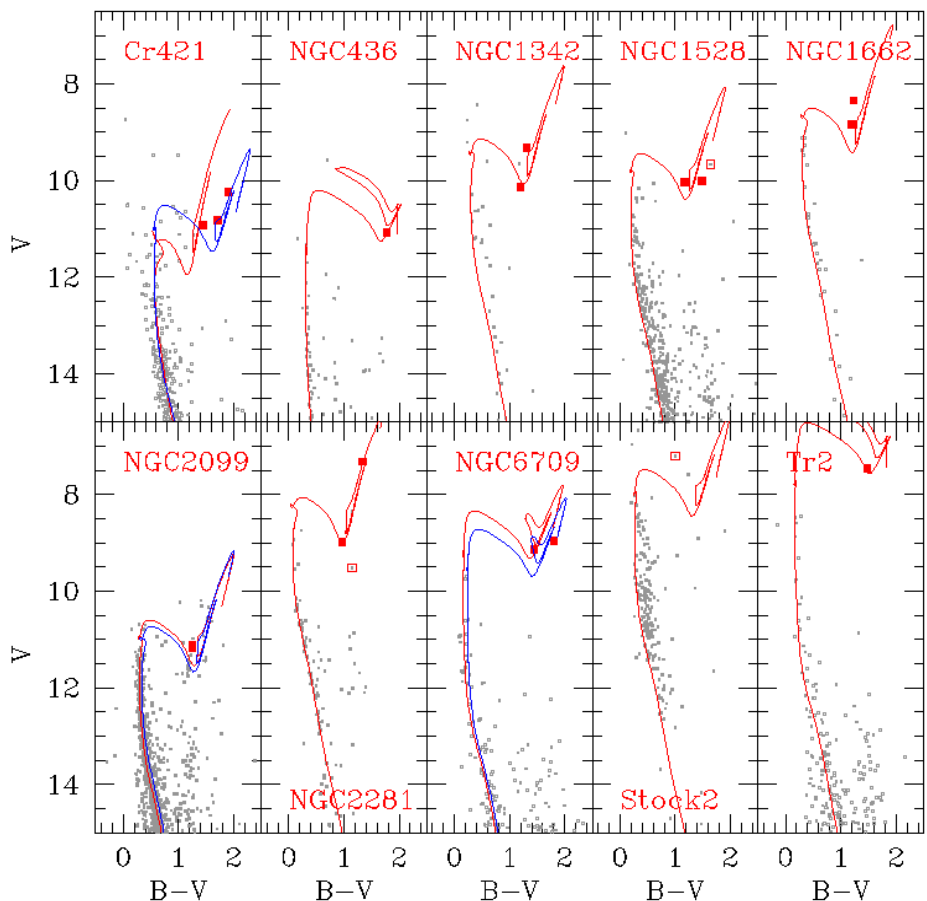}
 \caption{Colour-magnitude diagrams of the open clusters in our sample. The giants that were observed are shown as red filled squares (for cluster members) and red open squares (for the non-members). PARSEC isochrones \citep{2012MNRAS.427..127B} with solar metallicity are shown in each panel. The red isochrones were computed with cluster data from \citet{2013A&A...558A..53K}; the blue isochrones were computed with data from other sources (see references in Table \ref{tab:ocs}).}\label{fig:cmd}%
\end{figure*}
%

%-------------------------------------------------------------------

\section{Observational data}\label{sec:data}

The new spectra of open cluster giants were obtained with FIES \citep[FIber-fed Echelle Spectrograph;][]{2014AN....335...41T} at the 2.56 m Nordic Optical Telescope at Roque de los Muchachos Observatory, La Palma (Canary Islands, Spain). The FIES instrument is a cross-dispersed high-resolution echelle spectrograph. The spectra have resolution of R = 67\,000 and cover the spectral range between 370 nm and 730 nm. The observations were conducted in visitor mode during three nights between 5-8 September 2016. The data were reduced with the {\sf FIEStool} pipeline\footnote{http://www.not.iac.es/instruments/fies/fiestool/}. A log book of the observations is given in Table \ref{tab:log}. The list of clusters and their properties is given in Table \ref{tab:ocs}. Colour-magnitude diagrams (CMDs) of the open clusters are shown in Fig. \ref{fig:cmd}. We used PARSEC isochrones \citep{2012MNRAS.427..127B} to estimate the turn-off masses of the clusters. We chose these isochrones to be consistent with the choice made in \citet{2016A&A...589A.115S}. For the discussion of the results we join our new sample of open clusters with the sample presented in that paper.

In the subsections below, we discuss membership of the stars that were observed in each cluster. When possible, together with our radial velocities, we use proper motions from the HSOY \citep[Hot Stuff for One Year;][]{2017A&A...600L...4A} and UCAC5 \citep{2017AJ....153..166Z} catalogues. The proper motions are listed in Table \ref{tab:pm}. We also describe the Cepheids selected from the literature.

\begin{table*}
 \caption[]{\label{tab:log} Log book of the new spectroscopic observations of red giants in open clusters.}
\centering\small
\begin{tabular}{lccccccc}
\hline
\hline
Star &                R.A.        &         DEC      & $V$  & exp.\ time &      RV         & S/N             & Date of Obs. \\
        &  h:m:s (J2000) & d:m:s (J2000) & mag & s              & km s$^{-1}$ & @6700 \AA & (YYYY-MM-DD) \\
\hline
Cr 421 466   &   20:23:19.93 & +41:41:46.23 & 10.23$^{1}$ & 3 $\times$ 1000 & +1.8 & 210 & 2016-09-07 \\
Cr 421 529 &     20:23:28.73 & +41:41:23.91 & 10.68$^{1}$ & 3 $\times$ 1500 & +3.0 & 290 & 2016-09-06 \\
Cr 421 566 &     20:23:36.15 & +41:44:01.27 & 10.83$^{1}$ & 3 $\times$ 2400 & +1.8 & 220 & 2016-09-05 \\
NGC 436 482       &     01:15:57.74 & +58:47:57.30 & 11.07$^{2}$ & 3 $\times$ 2000 & $-$73.6 & 150 & 2016-09-07 \\
NGC 1342 4        &    03:32:11.23 & +37:22:55.45 & 9.33$^{3}$ & 3 $\times$ 500 & $-$10.5 & 100 & 2016-09-07 \\
NGC 1342 7        &  03:32:02.47 & +37:21:21.58 & 10.13$^{3}$ & 3 $\times$ 1000 & $-$10.3 & 190 & 2016-09-07 \\
NGC 1528 42       &     04:15:19.71 & +51:14:23.28 & 10.04$^{4}$ & 3 $\times$ 1000 & $-$10.6 & 170 & 2016-09-07 \\
NGC 1528 1009     &    04:14:55.21 & +51:10:42.29 & 10.00$^{4}$ & 3 $\times$ 500 & $-$9.5 & 110 & 2016-09-08 \\
$*$NGC 1528 4876     &    04:14:33.55 & +51:28:05.61 & 9.67$^{4}$ & 4 $\times$ 400 & +14.8 & 100 & 2016-09-08 \\
NGC 1662 1        &    04:48:29.51 & +10:55:48.27 & 8.34$^{3}$ & 3 $\times$ 300 & $-$13.5 & 190 & 2016-09-07 \\
NGC 1662 2        &    04:48:32.08 & +10:57:59.02 & 8.83$^{3}$ & 3 $\times$ 200 & $-$12.9 & 110 & 2016-09-06 \\
NGC 2099 1898     &   05:52:12.51 & +32:38:42.80 & 11.19$^{5}$ & 2 $\times$ 2000 & +9.0 & 150 & 2016-09-08 \\
NGC 2099 2520     &   05:52:16.54 & +32:34:45.90 & 11.12$^{5}$ & 2 $\times$ 1800 & +8.1 & 170 & 2016-09-08 \\
NGC 2281 55       &    06:48:15.10 & +41:04:22.20  & 8.98$^{6}$ & 3 $\times$ 450 & +19.6 & 120 & 2016-09-07 \\
NGC 2281 63       &    06:48:21.73 & +41:18:08.43 & 7.32$^{6}$ & 2 $\times$ 200 & +19.2 & 240 & 2016-09-07 \\
$*$NGC 2281 74       &    06:48:27.97 & +41:04:50.80  & 9.52$^{6}$ & 3 $\times$ 500 & $-$5.9 & 130 & 2016-09-08 \\
NGC 6709 208      &  18:51:31.80 & +10:18:49.87 & 9.13$^{7}$ & 3 $\times$ 400 & $-$10.4 & 120 & 2016-09-05 \\
NGC 6709 1998    &   18:51:11.20 & +10:18:18.03   & 8.95$^{7}$ & 3 $\times$ 400 & $-$20.8 & 160 & 2016-09-05 \\
$*$Stock 2 160       &    02:19:41.53 & +60:00:46.99 & 7.20$^{8}$ & 3 $\times$ 200 & +3.3 & 250 & 2016-09-06 \\
Trumpler 2 1      &    02:36:52.80 & +55:54:55.49 & 7.46$^{9}$ & 3 $\times$ 200 & $-$4.2 & 210 & 2016-09-06 \\
\hline
\end{tabular}
\tablefoot{The $V$ magnitudes are from (1) \citet{2007A&A...467.1065M}; (2) \citet{1994ApJS...90...31P}; (3) \citet{1961PUSNO..17..343H}; (4) \citet{2006AJ....132.1669S}; (5) \citet{2001AJ....122.3239K}; (6) \citet{1978PASJ...30..123Y}; (7) \citet{1999AJ....117..937S}; (8) \citet{1967ApJ...147..988K}; and (9) \citet{2006A&A...451..901F}. The radial velocities (RVs) are heliocentric and determined from our spectra. The coordinates for NGC 6709 1998 are from Donati, P., unpublished. ($*$) Stars marked with an asterisk were found to be non-members of the clusters (Section \ref{sec:data}).}
\end{table*}

\begin{table}
 \caption[]{\label{tab:ocs} Basic properties of the open clusters.}
\centering\small
\begin{tabular}{cccccccc}
\hline
\hline
Cluster &  distance &   E($B-V$)  & Age & M$_{\rm TO}$ \\
    & (pc) & (mag) & (Myr) & M$_{\odot}$ \\
\hline
\object{Collinder 421}  & 758   & 0.312  & 250$^{1}$  & 3.5 \\
\object{NGC 436}   &    3000 &  0.460 & 80 & 5.6 \\
\object{NGC 1342}  &     665 &  0.331  & 400 & 3.0 \\
\object{NGC 1528}  &     950  & 0.250  & 350 & 3.1 \\
\object{NGC 1662}  &     437  & 0.300  & 500 & 2.7 \\
\object{NGC 2099}  &    1400 &  0.350 & 430$^{2}$ & 2.9 \\
\object{NGC 2281}  &     500  & 0.065   & 610 & 2.5 \\
\object{NGC 6709}   &   1075 &  0.281 & 160$^{3}$ & 4.2 \\
\object{Stock 2}    &    400  & 0.339  & 280 & 3.4 \\
\object{Trumpler 2}  &   670  & 0.321 & 80 & 5.6 \\
\hline
\end{tabular}
\tablefoot{Distances and colour excess are from \citet{2013A&A...558A..53K}. Ages are mostly from same reference unless indicated: (1) \citet{2007A&A...467.1065M}; (2) \citet{2004MNRAS.351..649K}; (3) Donati, P., unpublished. The turn-off masses (M$_{\rm TO}$) are from PARSEC isochrones \citep{2012MNRAS.427..127B} of solar metallicity and the age of the cluster.}\end{table}

\subsection{Collinder 421}

Three stars were observed in the field of this cluster and we followed the numbering system of \citet{2007A&A...467.1065M}. The radial velocities and proper motions of the three stars are consistent within the uncertainties. \citet{2005A&A...438.1163K} gave membership probabilities of 64\%, 90\%, and 88\% for stars 466, 529, and 566, respectively. 

\begin{table*}
 \caption[]{\label{tab:pm} Proper motions for the cluster stars.}
\centering\small
\begin{tabular}{lccccccccc}
\hline
\hline
 Star      &   pmRA & $\sigma_{\rm pmRA}$ & pmDE & $\sigma_{\rm pmDE}$ & pmRA & $\sigma_{\rm pmRA}$ & pmDE & $\sigma_{\rm pmDE}$ \\ 
              &   mas/yr   &   mas/yr    &   mas/yr    &   mas/yr    &   mas/yr   &   mas/yr    &   mas/yr    &   mas/yr   \\
              &  (HSOY) & (HSOY) & (HSOY) & (HSOY) & (UCAC5) & (UCAC5) & (UCAC5) & (UCAC5) \\
\hline
Cr 421 466 &     $-$2.32  &   0.86 &   $-$8.58  &  0.85  &   $-$2.5     &  1.1    &   $-$8.7     &  1.1    \\ 
Cr 421 529 &     $-$3.53  &   0.84 &   $-$8.92  &  0.80  &   $-$3.6     &  1.0    &   $-$8.5     &  1.0    \\
Cr 421 566 &     $-$2.20  &   0.76 &   $-$7.29  &  0.72  &   $-$3.5     &  1.0    &   $-$8.5     &  1.0    \\
NGC 436 482       &    $-$3.48  &   1.30 &   $-$4.61  &  1.33  &   $-$1.6     &  1.4    &    +0.7     &  1.4    \\
NGC 1342 4        &   +0.42  &   0.04 &   $-$1.71  &  0.04  &    +0.3     &  1.8    &   $-$2.2     &  1.8    \\
NGC 1342 7        &         --   &    --   &     --    &   --    &     --     &   --    &     --     &   --    \\  
NGC 1528 42       &     +1.53  &   0.84 &   $-$0.67  &  0.94  &    +3.0     &  1.3    &   $-$2.4     &  1.3    \\
NGC 1528 1009     &      +0.02  &   0.95 &   $-$2.07  &  1.12  &    +2.1     &  1.5    &   $-$2.1     &  1.5    \\
$*$NGC 1528 4876     &     +2.75  &   0.71 &   $-$3.58  &  0.84  &    +3.6     &  1.5    &   $-$5.6     &  1.4    \\
NGC 1662 1        &     $-$1.04  &   0.35 &   $-$1.53  &  0.47  &    +1.6     &  2.3    &   $-$2.7     &  2.3    \\
NGC 1662 2        &     $-$1.19  &   0.04 &   $-$1.59  &  0.04  &   $-$1.9     &  1.7    &   $-$1.6     &  1.7    \\
NGC 2099 1898     &       +2.87  &   0.80 &   $-$5.84  &  0.84  &    +1.8     &  1.1    &   $-$5.5     &  1.1    \\
NGC 2099 2520     &     +3.15  &   0.91 &   $-$7.70  &  0.85  &    +1.3     &  1.1    &   $-$5.4     &  1.1    \\
NGC 2281 55       &    $-$3.17  &   0.48 &   $-$7.38  &  0.55  &    +0.8     &  1.4    &   $-$7.9     &  1.4    \\
NGC 2281 63       &     $-$3.02  &   0.04 &   $-$8.24  &  0.04  &   $-$1.8     &  1.7    &  $-$10.9     &  1.7    \\
$*$NGC 2281 74       &     $-$9.12  &   0.73 &  $-$19.48  &  0.76  &   $-$7.1     &  1.4    &  $-$19.4     &  1.4    \\
NGC 6709 208      &      +0.85  &   0.59 &   $-$3.72  &  0.63  &    +0.0     &  1.6    &   $-$4.6     &  1.6    \\
NGC 6709 1998     &         --   &    --   &     --    &   --    &     --     &   --    &     --     &   --    \\  
$*$Stock 2 160       &    $-$11.60  &   0.04 &   $-$8.56  &  0.04  &  $-$11.8     &  2.1    &   +30.8     &  2.1    \\
Trumpler 2 1      &    1.54  &   0.20 &   $-$5.53  &  0.20  &    +0.8     &  3.0    &    +4.6     &  3.0    \\
\hline
\end{tabular}
\tablefoot{Stars marked with an asterisk are considered to be non-members of the clusters.}
\end{table*}

\begin{table*}
\caption{Data of the selected sample of Cepheids.}
\label{tab:cepheids}
\centering\tiny
\begin{tabular}{cccccccccc} 
\hline\hline 
 Cepheid  & RA  & DEC & Galactocentric\tablefootmark{a} & Mass\tablefootmark{b} & [C/H]\tablefootmark{c} & [N/H]\tablefootmark{c} & [O/H]\tablefootmark{c} & [Na/H]\tablefootmark{d} & [Fe/H]\tablefootmark{d}\\
          &    h:m:s (J2000)      &     d:m:s (J2000)     &        distance (kpc)           &       M$_{\odot}$        &           dex          &           dex          &           dex          &            dex          &           dex          \\
\hline
\object{AQ Pup}    & 07:58:22.09  & $-$29:07:48.34  & 9.472$\pm$0.436 &  9.14$\pm$0.53 & $-$0.44 & 0.25 & $-$0.14 & 0.36$\pm$0.00 &  0.06$\pm$0.05 \\
\object{BB Sgr}    & 18:50:59.87  & $-$20:17:42.83  & 7.203$\pm$0.452 &  4.97$\pm$0.23 &   --    &    --  &   --    & 0.49          &  0.11          \\
\object{BF Oph}    & 17:06:05.50  & $-$26:34:50.03  & 7.081$\pm$0.452 &  3.42$\pm$0.17 & $-$0.13 & 0.44 &  0.21 & 0.41          &  0.13          \\
\object{BN Pup}    & 08:06:21.48  & $-$30:05:48.77  & 9.930$\pm$0.428 &  6.44$\pm$0.40 & $-$0.31 & 0.53 &  0.24 & 0.22$\pm$0.02 &  0.03$\pm$0.05 \\
\object{CV Mon}    & 06:37:04.85  & +03:03:50.28  & 9.362$\pm$0.452 &  4.08$\pm$0.21 & $-$0.30 & 0.20 &  0.05 & 0.31$\pm$0.20 &  0.09$\pm$0.09 \\
\object{$\delta$ Cep} & 22:29:10.27 & +58:24:54.71 & 8.010$\pm$0.452 &  4.53$\pm$0.26 & $-$0.17 & 0.36 &  0.23 & 0.38          &  0.11          \\
\object{$\eta$ Aql}   & 19:52:28.37 & +01:00:20.37 & 7.750$\pm$0.452 &  4.36$\pm$0.25 &   --    &    --  &     --  & 0.33          &  0.11          \\
\object{EV Sct}    & 18:36:39.60  & $-$08:11:05.36  &      --        &  5.46$\pm$0.60 & $-$0.23 & 0.32 &  0.23 & 0.25$\pm$0.00 &  0.09$\pm$0.07 \\
\object{KN Cen}    & 13:36:36.90  & $-$64:33:29.92  &      --        & 12.45$\pm$0.72 & $-$0.04 & 0.42 &  0.33 & 0.69$\pm$0.04 &  0.55$\pm$0.12 \\
\object{l Car}  & 09:45:14.81 & $-$62:30:28.45 & 7.845$\pm$0.451 & 13.71$\pm$0.66 & $-$0.11 & 0.04 &  0.22 & 0.09          &  0.17          \\
\object{RS Pup}    & 08:13:04.22  & $-$34:34:42.70  & 8.585$\pm$0.444 & 12.79$\pm$0.94 & $-$0.54 & 0.54 &  0.56 & 0.75          &  0.14          \\
\object{RY Sco}    & 17:50:52.34 & $-$33:42:20.42 & 6.663$\pm$0.453 &  6.22$\pm$0.32 &   --    & --     & --      & 0.36$\pm$0.06 &  0.01$\pm$0.06 \\
\object{RY Vel}    & 10:20:41.03  & $-$55:19:17.03  & 7.774$\pm$0.432 &  8.51$\pm$0.41 & $-$0.05 & 0.40 &  0.29 & 0.46          &  0.02          \\
\object{RZ Vel}    & 08:37:01.30  & $-$44:06:52.85  & 8.249$\pm$0.445 &  7.89$\pm$0.38 & $-$0.33 & 0.18 &  0.09 & 0.45          &  0.12          \\
\object{S Nor}    & 16:18:51.83  & $-$57:53:59.25  & 7.232$\pm$0.451 &  6.64$\pm$0.32 & $-$0.12 & 0.38 &  0.21 & 0.45          &  0.06          \\
\object{SU Cas}    & 02:51:58.75  & +68:53:18.60  & 8.229$\pm$0.451 &  6.56$\pm$0.51 &    --   &    --  &   --    & 0.39          &  0.09          \\
\object{SV Vul}    & 19:51:30.91  & +27:27:36.84  & 7.284$\pm$0.438 & 13.93$\pm$0.74 &   --    &  --    &   --    & 0.18          &  0.08          \\
\object{SW Vel}    & 08:43:38.69  & $-$47:24:11.20  & 8.457$\pm$0.433 &  7.35$\pm$0.34 & $-$0.46 & 0.49 &  0.04 & 0.23          & $-$0.07          \\
 \object{T Mon}    & 06:25:13.00 & +07:05:08.56 & 9.056$\pm$0.452 & 10.23$\pm$0.68 & $-$0.15 & 0.47 &  0.33 & 0.53          &  0.00          \\
 \object{T Vel}    & 08:37:40.82  & $-$47:21:43.08  & 8.084$\pm$0.448 &  3.30$\pm$0.23 & $-$0.35 & 0.17 &  0.07 & 0.35          &  0.05          \\
 \object{U Car}    & 10:57:48.19  & $-$59:43:55.88  & 7.574$\pm$0.443 &  8.15$\pm$0.39 &  0.17 & 0.42 &  0.07 & 0.53          &  0.21          \\
 \object{U Nor}    & 15:42:20.92  & $-$55:18:42.94  & 6.876$\pm$0.450 &  5.90$\pm$0.41 &  0.09 & 0.56 &  0.40 & 0.41          &  0.11          \\
 \object{U Sgr}    & 18:31:53.33  & $-$19:07:30.26  & 7.371$\pm$0.452 &  4.50$\pm$0.20 &     --  &   --   &   --    & 0.34          &  0.11          \\
\object{UU Mus}    & 11:52:17.73  & $-$65:24:15.06  & 7.097$\pm$0.414 &  6.00$\pm$0.54 & $-$0.22 & 0.37 &  0.12 & 0.37          &  0.15          \\
\object{V340 Nor}  & 16:13:17.68   & $-$54:14:05.06   &     --         &  5.09$\pm$0.23 & $-$0.08 & 0.34 &  0.27 & 0.40$\pm$0.00 &  0.07$\pm$0.07 \\
 \object{V Cen}    & 14:32:33.08  & $-$56:53:15.78  & 7.459$\pm$0.451 &  4.35$\pm$0.31 & $-$0.22 & 0.30 &  0.11 & 0.38          &  0.08          \\
\object{VW Cen}    & 13:33:59.03  & $-$64:03:20.00  &     --         &  6.11$\pm$0.32 &   --    &  --    &   --    & 0.79$\pm$0.05 &  0.41$\pm$0.08 \\
\object{VY Car}    & 10:44:32.70  & $-$57:33:55.33  & 7.627$\pm$0.441 &  8.47$\pm$0.39 & $-$0.21 & 0.59 &  0.17 & 0.31          &  0.01          \\
\object{WZ Car}    & 10:55:18.73  & $-$60:56:23.95  & 7.601$\pm$0.396 &  6.47$\pm$0.49 & $-$0.40 & 0.11 & $-$0.09 & 0.53          &  0.02          \\
\object{WZ Sgr}    & 18:16:59.72 & $-$19:04:32.97 & 6.326$\pm$0.453 &  8.22$\pm$0.49 &    --   &  --    &   --    & 0.59$\pm$0.07 &  0.28$\pm$0.08 \\
 \object{X Cyg}    & 20:43:24.19  & +35:35:16.07  & 7.772$\pm$0.448 &  8.61$\pm$0.40 &     --  &  --    &  --     & 0.38          &  0.13          \\
\object{XX Cen}    & 20:43:24.19  & +35:35:16.07  & 7.015$\pm$0.445 &  5.69$\pm$0.26 & $-$0.17 & 0.39 &  0.19 & 0.34          &  0.06          \\
\hline           
\end{tabular}
\tablefoot{
\tablefoottext{a}{Distances from \citet{2014A&A...566A..37G}. The distances were computed from near-IR period-Wesenheit relations \citep{2013ApJ...764...84I}}
\tablefoottext{b}{Pulsation masses of Milky Way Cepheids derived using predicted period-luminosity-colour relations with Z = 0.02 and Y = 0.25-0.31 \citep{2005ApJ...629.1021C}}
\tablefoottext{c}{From \citet{2011AJ....142..136L}
}
\tablefoottext{d}{From \citet{2015A&A...580A..17G}
}
}
\end{table*}

\subsubsection{Star Cr 421 566: Fast rotator with infrared excess}\label{sec:lirich}

Star 566 is a moderately fast rotator (we estimate $v~\sin~i$ $\sim$ 18 km s$^{-1}$). This star was found to have infrared (IR) excess by \citet{2005MNRAS.363.1111C}, using data from the Mid-Course Space Experiment (MSX) Point Source Catalogue \citep{2001AJ....121.2819P}. The measured excess was E($Ks-[8]$)\footnote{$Ks$ is from 2MASS \citep{2006AJ....131.1163S} and [8] is a converted magnitude from the MSX flux at 8 $\mu$m.} = 0.55 mag. Stars with excess at $Ks-[8]$ of the order of 1 mag were associated by \citet{2005MNRAS.363.1111C} with asymptotic giant branch stars with enhanced mass loss. Based on this, the observed IR excess of star 566 might be seen as a sign of enhanced mass loss.

Lithium-rich giants \citep[e.g.][]{2016MNRAS.461.3336C} seem to be common among fast rotators \citep{2002AJ....123.2703D} and to have IR excess sometimes \citep{2015AJ....150..123R,2015A&A...577A..10B}, although the excess is usually detected at about 20 $\mu$m. The IR excess might be connected to enhanced mass loss \citep{1996ApJ...456L.115D,2015ApJ...806...86D}. Infrared data for star 566 is however not available at wavelengths longer than 8 $\mu$m. We thus decided to check whether star 566 would turn out to be a new Li-rich giant. 

We computed synthetic spectrum around the Li 6708 \AA\ line using {\sf MOOG}\footnote{\url{http://www.as.utexas.edu/~chris/moog.html}} \citep{1973PhDT.......180S,2012ascl.soft02009S} and combining the atomic line list of \citet{2012A&A...543A..29M} with the molecular line list described below (Sect. \ref{sec:cno}). The Li line was detectable with A(Li) $\sim$ 1.1 $\pm$ 0.1 in LTE, where the error takes into account only the uncertainty of the fit itself. While the derived Li abundance is somewhat uncertain, given the rotational broadening, we can definitely conclude that star 566 is not a Li-rich giant. 

\subsection{NGC 436}

One star was observed in this cluster, NGC 436 482 in the numbering system of \citet{1994AJ....107.1079P}. \citet{2008A&A...485..303M} found the star to have a mean RV = $-$74.0 $\pm$ 0.14 (0.21 rms) km s$^{-1}$, which is in agreement with our determination. \citet{2014A&A...564A..79D} found the star to have 99\% membership probability based on UCAC4 proper motions \citep{2013AJ....145...44Z}.

\subsection{NGC 1342}

Stars 4 and 7 in the system of \citet{1961PUSNO..17..343H} were observed. \citet{2015MNRAS.450.4301R} determined stars 4 and 7 to have RV = $-$10.9 and $-$10.8 km s$^{-1}$, respectively, in agreement with our values. \citet{2008A&A...485..303M} found RVs = $-$10.9 and $-$10.7 km s$^{-1}$ for the stars 4 and 7, respectively, which again agree with our determination. \citet{2014A&A...564A..79D} found both stars to have 98\% membership probability based on UCAC4 proper motions.%  

\subsection{NGC 1528}

Stars 42, 1009, and 4876 in the numbering system of \citet{2006AJ....132.1669S} were observed. \citet{2008A&A...485..303M} found a mean RV = $-$10.52 and $-$9.85 kms$^{-1}$ for stars 42 and 1009, respectively, which is in agreement with our values. Star 42 was flagged by \citet{2008A&A...485..303M} as a possible spectroscopic binary, but the fact that the RV values agree argues against that. Star 4876 was found to have a very different RV, +14.8 km s$^{-1}$, and we consider this object a cluster non-member. \citet{2014A&A...564A..79D} derived 99\% and 90\% membership probability for stars 42 and 1009. The HSOY and UCAC5 proper motions for the same star tend to differ (Table \ref{tab:pm}). There is also some scatter in these values among the stars in one given catalogue. Within the uncertainties, however, the values are still somewhat consistent. 

\subsection{NGC 1662}

Stars 1 and 2, according to the system of \citet{1961PUSNO..17..343H}, were observed. \citet{2015MNRAS.450.4301R} derived RVs in agreement with ours of $-$13.6 and $-$12.9 km s$^{-1}$ for stars 1 and 2, respectively. \citet{2008A&A...485..303M} found mean RV = $-$13.9 and $-$13.0 km s$^{-1}$ for stars 1 and 2, respectively, which is also in agreement with our values. \citet{2014A&A...564A..79D} found both stars to have 99\% membership probability. The HSOY and UCAC5 proper motions agree within the uncertainties. 

Star 1 is listed in CCDM \citep[Catalog of Components of Double and Multiple stars;][]{2002yCat.1274....0D} and WDS \citep[Washington Double Star catalogue;][]{2001AJ....122.3466M} as part of a system of five components (named HJ 684). \citet{2013AJ....146...56M} found separations of 24.2$\arcsec$ between components A and B and of 77.6$\arcsec$ between components A and E using speckle interferometry. However, the stable RV values of star 1 as determined in various works argue that this is likely a single star.

\subsection{NGC 2099}

Stars 1898 and 2520, which are stars 24 and 20 in the numbering system of \citet{2001AJ....122.3239K}, were observed. \citet{2010A&A...511A..56P} derived RV = +8.79 km s$^{-1}$ for star 2520 (their star 67), which is in reasonable agreement with our value. \citet{2008A&A...485..303M} found a mean RV = +8.78 and +8.04 km s$^{-1}$ for stars 1898 and 2520, respectively, which is also in agreement with our values. \citet{2014A&A...564A..79D} found the stars 1898 and 2520 to have 99\% and 98\% membership probability, respectively. The HSOY and UCAC5 proper motions agree within the uncertainties. 
\subsection{NGC 2281}

Three stars were observed in NGC 2281: stars 55, 63, 74 in the numbering system of \citet{1959AJ.....64..170V}. \citet{2008A&A...485..303M} found mean RV = +19.1 and +19.0 km s$^{-1}$ for stars 55 and 63, respectively, in good agreement with our values. \citet{2005A&A...430..165F} found RV = +18.96 km s$^{-1}$ for star 63, again in agreement with the value found here. 

\citet{2014A&A...564A..79D} found the stars 55, 63, and 74 to have 99\%, 98\% and 0\% membership probability, respectively. The HSOY and UCAC5 proper motions of star 74 are very different from those of the two other stars. It also has a different RV ($-$5.9 km s$^{-1}$). Therefore, star 74 is likely not a member of the cluster.

\subsection{NGC 6709}

Two stars were observed, numbers 208 and 1998 in the system adopted within WEBDA\footnote{An on-line database devoted to stellar clusters in the Galaxy and Magellanic clouds, see \url{http://webda.physics.muni.cz}}. Star 208 also has number 208 in \citet{1983A&AS...51..541H}. \citet{2008A&A...485..303M} found mean RV = $-$10.68 km s$^{-1}$ for star 208, which is in good agreement with our value. \citet{2014A&A...564A..79D} found star 208 to have 97\% membership probability. No membership study including star 1998 seems to be available.

Star 208 is listed in CCDM and WDS as component C in a quadruple system (named HJ 870). The separation between components A and C (star 208) is of 65.8$\arcsec$, and between components C and D of 22$\arcsec$ \citep{2001AJ....122.3466M}. This multiplicity seems incompatible with the stable RV measurements.

\setcounter{table}{6}
\begin{table}
 \caption[]{\label{tab:cnosigma} Uncertainties in the abundances of C, N, and O.}
\centering\small
\begin{tabular}{lcccc}
\hline
\hline
Parameter  &  Change & $\Delta_{\rm A(C)}$ & $\Delta_{\rm A(N)}$ & $\Delta_{\rm A(O)}$ \\
\hline  
$T_{\rm eff}$ (K)   & +100    &  0.00    & +0.03    & +0.02 \\
$T_{\rm eff}$ (K)   &  $-$100  &   $-$0.01 &    +0.02  &   $-$0.01 \\
$\log~g$ (dex) &  +0.15   & +0.04   &  +0.06   &  +0.07 \\
$\log~g$ (dex)  & $-$0.15  &  $-$0.05  &   $-$0.03  &   $-$0.06 \\
{[Fe/H]} (dex)   &   +0.15  &  +0.05   &  +0.09  &   +0.06 \\
{[Fe/H]} (dex)    & $-$0.15  &  $-$0.06   &  $-$0.06   &  $-$0.06 \\
$\xi$ (km/s)  & +0.10  &   0.00   &  +0.01  &   +0.01 \\
$\xi$ (km/s)  & $-$0.10  &  $-$0.01   &  +0.01  &   $-$0.01 \\
\hline
Total uncertainty & -- & $\pm$0.08 & $\pm$0.11 & $\pm$0.09 \\
\hline
\end{tabular}
%\tablefoot{}
\end{table}

\subsection{Stock 2}

Star number 160 \citep{1967ApJ...147..988K} was observed towards Stock 2. The proper motions from \citet{2000A&AS..146..251B}, however, indicate that the star is not a member of the cluster. \citet{2008A&A...485..303M} found a mean RV = 3.42 km s$^{-1}$, which is in good agreement with our value. They also concluded that the star is not a member of the cluster. \citet{1999A&AS..139..433D} found RV = 2.98 $\pm$ 0.12 km s$^{-1}$, a smaller value but still consistent. We thus have to consider that the star is likely not a member of the cluster.

\subsection{Trumpler 2}

Star number 1 in the system of \citet{1961PUSNO..17..343H} was observed. \citet{2008A&A...485..303M} found a mean RV = $-$3.16 km s$^{-1}$, which is somewhat different from our value of $-$4.2 km s$^{-1}$. \citet{2014A&A...564A..79D} found the star to have 97\% membership probability. We consider the star to be a cluster member. A companion was observed with speckle interferometry by \citet{2012AJ....143..124M} with 35$\arcsec$ of mean separation. The WDS includes the star as part of system of five components (WDS 02369+5555) with separation ranging from 35$\arcsec$ to 138.6$\arcsec$. Although multiplicity can not be discarded, given the different RV measurements, it is perhaps unlikely that the star is part of a quintuple system. 
\subsection{Cepheids}\label{sec:cepheids}

In order to extend the mass range towards higher masses, we included in this study data of 32 Galactic Cepheids. The masses of these Cepheids range from 3 $M_{\odot}$ to 14 $M_{\odot}$, the bulk of the sample having masses between 4 and 10 $M_{\odot}$. We used masses computed with the help of predicted period-luminosity-colour (PLC) relations in the $V$ and $K$ bands \citep{2005ApJ...629.1021C}. The PLC relations were computed with a metal content of Z=0.02 and their intrinsic dispersion includes the variation of the helium content from Y=0.25 to Y=0.31. 

Until recently, pulsation and evolutionary models of Cepheids led to discrepant results concerning the determination of their mass: masses derived from evolutionary models were $\sim$ 20\% higher than masses derived from pulsation models. The presence of a Cepheid in a double-lined eclipsing binary (in the Large Magellanic Cloud) enabled \citet{2010Natur.468..542P} to determine its mass with an unprecedented precision, strongly supporting masses derived using the pulsation theory. The Cepheids and their properties are listed in Table \ref{tab:cepheids}.

%-------------------------------------------------------------------
\section{Spectroscopic analysis}\label{sec:analysis}

\subsection{Open cluster giants}

\subsubsection{Atmospheric parameters and abundances}

Our analysis is based on equivalent widths (EWs) measured using the {\sf DOOp} program \citep[DAOSPEC Option Optimiser;][]{2014A&A...562A..10C}, a wrapper  of {\sf DAOSPEC} \citep{2008PASP..120.1332S,2010ascl.soft11002S}. The observed RV is an output of DAOSPEC. Heliocentric corrections were applied and the resulting heliocentric RVs are listed in Table \ref{tab:log}.

The stellar atmospheric parameters (Table 5) were derived with the program {\sf FAMA} \citep[Fast Automatic MOOG Analysis;][]{2013A&A...558A..38M}, a wrapper of {\sf MOOG}.  We assumed LTE and used the MARCS model atmospheres \citep{2008A&A...486..951G}. The atomic data are the public version of those prepared for the \emph{Gaia}-ESO Survey \citep{2015PhyS...90e4010H} and based on VALD3 data \citep{2011KIzKU.153...61R}. Thus, the analysis inherits from part of the methods used in the spectral analysis of the \emph{Gaia}-ESO Survey \citep[see the description of the EPINARBO node in][]{2014A&A...570A.122S}, although we do not benefit from the multiple pipelines approach.

{\sf The FAMA} program uses the classical spectroscopic method to estimate the stellar parameters: {\it i)} $T_{\rm eff}$ is obtained by eliminating trends between the line abundances of a chemical element and the excitation potentials; {\it ii)} $\log~g$ is optimised by assuming the ionisation equilibrium, i.e. requiring that for a given species, the same abundance (within the uncertainties) is obtained from lines of two ionisation states (using, typically, neutral and singly ionised lines); and {\it iii)} $\xi$ is set by minimising the slope of the relationship between line abundances and the logarithm of the reduced EWs. We typically used about 200 \ion{Fe}{i} lines and 20 \ion{Fe}{ii} lines for the determination of stellar parameters.

The uncertainties in the atmospheric parameters (Table 5) reflect the uncertainty in the slopes used to constrain the parameters. The uncertainties in the abundances reflect the scatter among the various lines that were used. Abundances and uncertainties are given in Table \ref{tab:abun} in the Appendix. We adopt the solar abundance values from \citet{2007SSRv..130..105G} as reference.

\subsubsection{Abundances of C, N, and O using spectrum synthesis}\label{sec:cno}

The abundances of C, N, and O were determined with spectrum synthesis with the code {\sf MOOG}. The results are given in Table 6. The analysis is conducted assuming LTE and used the MARCS model atmospheres \citep{2008A&A...486..951G}. 

The C abundances were derived using the C$_{2}$ features around 5135 \AA, the N abundances using the CN band at 6332 \AA, and the O abundances using the forbidden [\ion{O}{i}] line at 6300 \AA. The synthesis of the O line took into account the \ion{Ni}{i} line at 6300.34 \AA\ \citep[with data from][]{2003ApJ...584L.107J} and the nearby \ion{Sc}{ii} line at 6300.70 \AA\ with hyperfine structure from \citet{1989A&A...222...35S}.

The molecular data for the synthesis include the C$_{2}$ Swan system \citep[$^{12}$C$^{12}$C and $^{12}$C$^{13}$C,][]{2013JQSRT.124...11B,2014ApJS..211....5R}, the CN red and violet systems \citep[$^{12}$C$^{14}$N, $^{13}$C$^{14}$N, and $^{12}$C$^{15}$N,][]{2014ApJS..214...26S}, and MgH \citep[$^{24}$MgH, $^{25}$MgH and $^{26}$MgH,][]{2013ApJS..207...26H}, which contaminates the region of the C$_{2}$ lines.

Abundances are not provided for Collinder 421 566, found to be a fast rotator, and for NGC1342 7 and NGC1662 2, which are relatively warm and thus have weak molecular features that result in very uncertain abundances. Uncertainties affecting the CNO abundances (Table \ref{tab:cnosigma}) were estimated with star NGC 6709 208 by changing each atmospheric parameter in turn, by its own uncertainty, and recomputing the abundances.

\subsubsection{Non-LTE corrections for the Na and Al abundances}

The surface Na abundances were corrected for non-LTE effects using the grids of \citet{2011A&A...528A.103L}\footnote{Using an interpolating routine in {\sf IDL} made available by Karin Lind (2011, private communication).}. The corrections were derived on a line-by-line basis using the atmospheric parameters and the LTE Na abundance of each star as input. Four Na lines were used to compute the abundances: 4751.822, 4982.814, 6154.225, and 6160.747 \AA. The computed corrections for the giants in open clusters range between $-$0.07 dex and $-$0.14 dex. The corrections have a mean value of $-$0.10 dex. The Na abundances listed in Table \ref{tab:abun} are given already corrected for the non-LTE effects.

Comparisons between Na abundances in giants derived using 1D and 3D model atmospheres were presented by \citet{2007A&A...469..687C} and \citet{2013A&A...559A.102D}. These studies have found that, for abundances based on the Na lines 6154 and 6160 \AA,~ the corrections are small ($\leq$ $\pm$0.05 dex) and could be positive. The Na abundances were not corrected for these 3D effects. The reference solar abundance adopted here, $\log \epsilon$(Na) = 6.17 \citep{2007SSRv..130..105G}, already includes non-LTE and 3D corrections. 

New grids of non-LTE corrections for Al abundances have recently been published by \citet{2017A&A...607A..75N}. We interpolated non-LTE corrections for the atmospheric parameters of our sample using the grid computed by \citet{2017A&A...607A..75N} with 1D MARCS photospheric models\footnote{Using an interpolating routine in {\sf IDL} made available by Thomas Nordlander (2017, private communication).}. As with Na, the corrections were derived on a line-by-line basis using as input the LTE Al abundance of each star. Two lines were used, i.e. 6696.021 and 6698.671 \AA. 

In addition, we computed and applied non-LTE corrections for the Al abundances of the giants discussed in \citet{2016A&A...589A.115S}, \citet{2017A&A...598A..68O}, and \citet{2017A&A...601A..56T}. As a consequence we have non-LTE Al abundances for giants covering the mass range between 1 M$_{\odot}$ and 6 M$_{\odot}$. The typical correction is of the order of $-$0.05 dex, varying between $-$0.01 and $-$0.09 dex. This work is the first to make use of this grid for a systematic investigation of non-LTE Al abundances in red giants of open clusters. The reference solar abundance of Al is $\log \epsilon$(Al) = 6.37 \citep{2007SSRv..130..105G}.

\subsection{Atmospheric parameters and abundances of the Cepheids}

Since they are pulsating stars, the atmospheric parameters of Cepheids vary over the period. Moreover, they have to be derived directly from the spectra because simultaneous photometry is generally not available. The $T_{\rm eff}$ is then determined from calibrations using the line-depth-ratio method \citep{2000A&A...358..587K}. A classical spectroscopic analysis (i.e. the ionisation and excitation equilibria of Fe lines) is the next step in the spectroscopic determination of the parameters.

For the Cepheids in our sample, we adopted the CNO abundances derived by \citet{2011AJ....142..136L} using atomic lines of these elements. For C and O, different weights based on relative strength and blending were attributed to different lines depending on $T_{\rm eff}$. The analysis assumed LTE.

Oxygen abundances from \citet{2011AJ....142..136L} are based both on the forbidden [\ion{O}{i}] line at 6300 \AA\ and the \ion{O}{i} triplet at 7775 \AA. While the forbidden line forms under LTE, the triplet is well known to be affected by non-LTE \citep[see e.g.][]{1993A&A...275..269K,1998PASJ...50...97T,2015A&A...583A..57S}.

In addition, non-LTE effects can be important for abundances derived from atomic lines of C and N \citep[e.g.][]{2011MNRAS.410.1774L,2015MNRAS.446.3447L}. For carbon, non-LTE corrections tend to be smaller for lines in the visual (e.g. 5052 and 5380 \AA) and increase for lines of longer wavelengths. Departures from LTE also increase for warmer and brighter stars. For nitrogen, non-LTE effects can be of the order of $\sim$+0.1 dex for stars with $T_{\rm eff}$ $\sim$ 5700 K and increase to $\sim$+0.8 dex at $T_{\rm eff}$ $\sim$ 8500 K.

For Na, we used abundances based on the lines at 6154.225 and 6160.747 \AA~\citep{2015A&A...580A..17G}. \citet{2015A&A...580A..17G} also compiled previous results from \citet{2006AJ....131.2256Y}, \citet{2011AJ....142...51L}, \citet{2011AJ....142..136L}, and \citet{2013A&A...558A..31L} and rescaled these results by $\sim$ $-$0.10 dex to take into account zero-point differences between different studies, thus providing a large, homogeneous sample of 439 Cepheid abundances. From this large sample, we selected only the stars with accurate masses as described above (Section \ref{sec:cepheids}). 

In the Sections below, the Na abundances of the Cepheids are discussed in LTE. Correcting these literature values for non-LTE effects is not straightforward. The abundances tend to be average values of multiple determinations from spectra taken at different phases, and thus where the Cepheid was characterised by different atmospheric parameters. Moreover, the grid of non-LTE corrections of \citet{2011A&A...528A.103L} only extends down to $\log~g$ = 1.0 dex (and at this $\log~g$ only to $T_{\rm eff}$ = 5500 K). Therefore, many brighter Cepheids fall outside the grid. To have an idea of the order of magnitude of the non-LTE corrections, we interpolated values for a star with $\log~g$ = 1.0 dex, $T_{\rm eff}$ = 5500 K, solar metallicity, and LTE abundance of [Na/Fe] = +0.30 dex using the grid of \citet{2011A&A...528A.103L}. The average value of the corrections for lines 6154 and 6160 \AA\ is about $-$0.06 dex.

\begin{table}
 \caption[]{\label{tab:tomasses} Comparison between the predicted turn-off masses for given ages in various stellar models.}
\centering\small
\begin{tabular}{cccc}
\hline
\hline
Age (Myr) & M$_{\rm TO}$ (M$_{\odot}$) & M$_{\rm TO}$ (M$_{\odot}$) & M$_{\rm TO}$ (M$_{\odot}$) \\
  & \citeauthor{2012MNRAS.427..127B} & {\citeauthor{2012A&A...537A.146E}}  & {\citeauthor{2012A&A...537A.146E}} \\
  & non-rotating & non-rotating & with rotation \\
\hline  
80 & 5.6 & 5.4 & 6.0 \\
160 & 4.2 & 4.0 & 4.4 \\
250 & 3.5 & 3.4 & 3.7 \\
280 & 3.4 & 3.4 & 3.7 \\
350 & 3.1 & 3.0 & 3.4 \\
400 & 3.0 & 2.8 & 3.0 \\
430 & 2.9 & 2.8 & 3.0 \\
500 & 2.7 & 2.6 & 2.8 \\
610 & 2.5 & 2.6 & 2.4 \\
\hline
\end{tabular}
\tablefoot{The \citet{2012MNRAS.427..127B} isochrones were computed for the exact age using the PARSEC on-line interface\footnote{\url{http://stev.oapd.inaf.it/cgi-bin/cmd}}. The isochrones from \citet{2012A&A...537A.146E} are available at steps of 0.1 dex in $\log({\rm Age})$. We list the turn-off values for these pre-computed isochrones with age closest to the age listed in the first column of the Table. }
\end{table}

\section{Discussion}\label{sec:disc}

In this Section, we compare the observed abundances as a function of stellar mass to predictions from various models of stellar evolution. In fact, the turn-off masses of the clusters are used as reference in these comparisons. Following a remark from the referee, we first compare the turn-off masses from the different models to understand whether this could have any impact in the discussion. Among the stellar models we use, those from \citet{2012A&A...543A.108L} and \citet{2013MNRAS.431.3642V} do not provide isochrones and thus are not part of this comparison. Table \ref{tab:tomasses} list the turn-off masses of isochrones with certain ages in the models of \citet{2012MNRAS.427..127B} and \citet{2012A&A...537A.146E}.

For the same age, the non-rotating models of \citet{2012A&A...537A.146E} predicted systematically lower turn-off masses than the models of \citet{2012MNRAS.427..127B}. The difference however is small, of at most 0.2 M$_{\odot}$, which is of the same magnitude as the error we assumed for the mass values. The biggest difference is not in the comparison between non-rotating models of different authors, but when comparing the \citeauthor{2012A&A...537A.146E} isochrones with and without rotation. Rotation increases the time that stars spend in the main sequence and the effect is larger for higher masses. For a given age, the turn-off mass is larger if rotation is included. Between 2-3 M$_{\odot}$, the difference between rotating and non-rotating models is of about 0.2 M$_{\odot}$, but this increases to 0.6 M$_{\odot}$ for mass values between 5-6 M$_{\odot}$. 

As shown in the discussion and figures below, these differences are small enough that, in their respective mass regimes, the conclusions are not affected. The trends between abundances and mass would be the same if different scales of turn-off masses were adopted.

\subsection{CNO abundances}

\begin{figure*}
\centering
\includegraphics[height = 7cm]{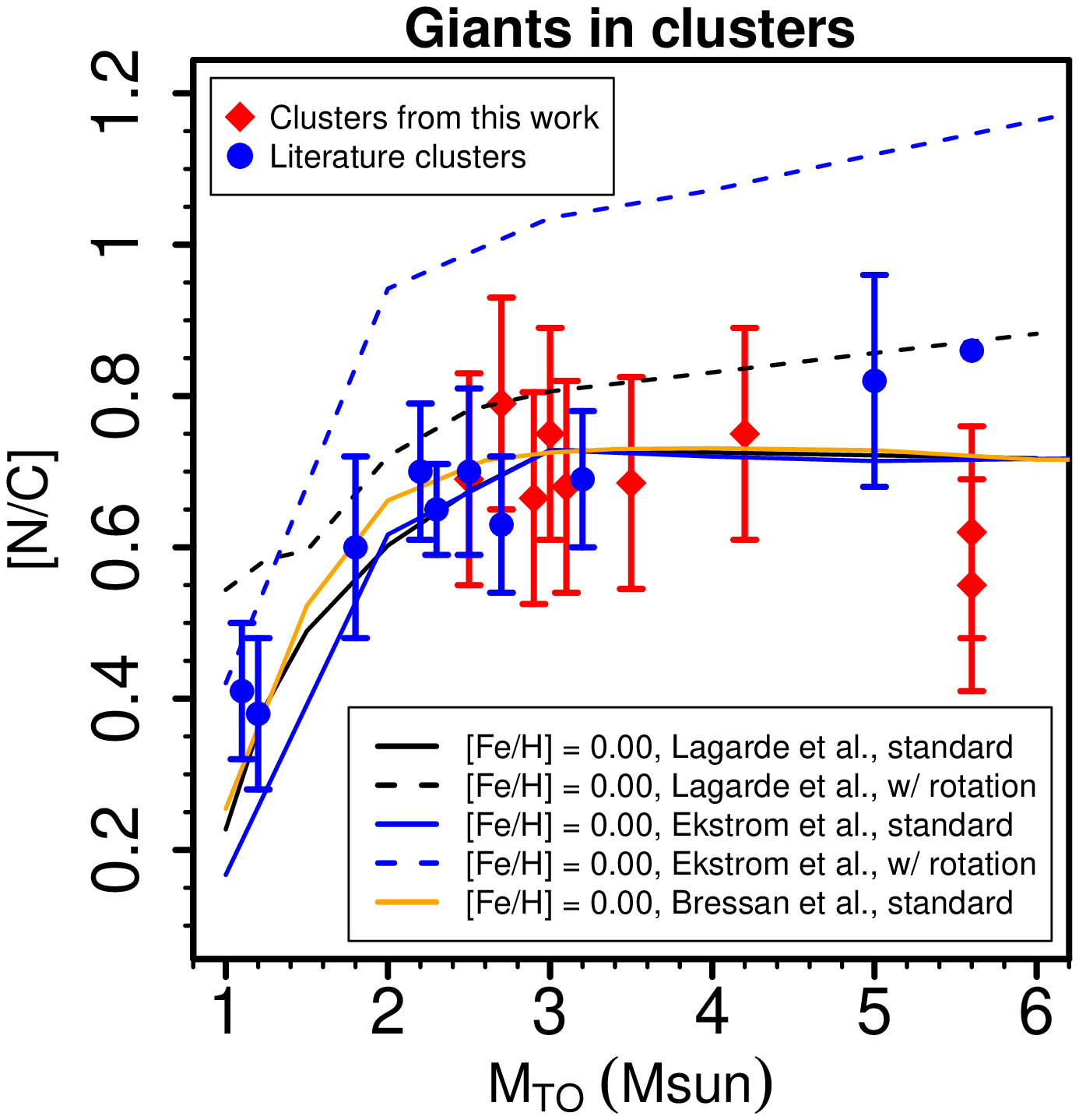}
\includegraphics[height = 7cm]{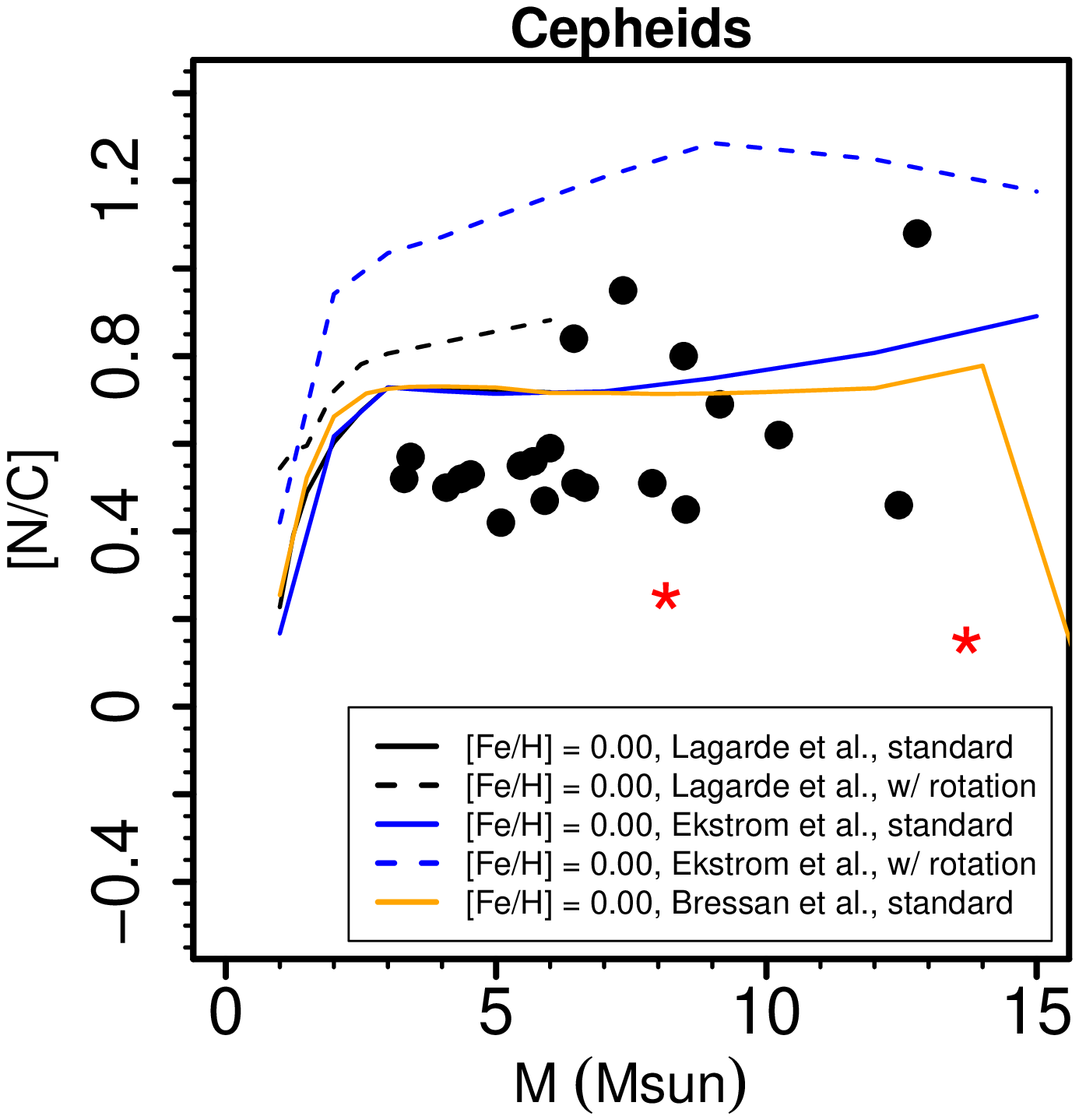}
 \caption{\emph{Left:} Mean [N/C] ratio of the clusters analysed in this work (red diamonds) computed using only  the member stars. The C and N abundances flagged as uncertain in Table 6 are not shown. The error bars depict the typical uncertainty in [N/C] propagated from the uncertainties in C and N listed in Table \ref{tab:cnosigma} and not the cluster scatter. Most clusters have good C and N abundances for one giant alone. A few selected results from the literature are included as blue symbols \citep{2015A&A...573A..55T,2016A&A...595A..16T,2016A&A...589A..50D,2016MNRAS.462..794D}. \emph{Right:} [N/C] ratio, in LTE, of selected Cepheids. References are given in Table \ref{tab:cepheids}. Two Cepheids suspected to be at the first crossing are shown as red stars; the other Cepheids are indicated as solid circles. The models of \citet[][]{2012A&A...543A.108L}, shown as black lines, extend only up to 6 M$_{\odot}$. The models of \citet{2012A&A...537A.146E} are shown as blue lines and those from \citet{2012MNRAS.427..127B} as an orange line.}\label{fig:cno}%
\end{figure*}

The left panel of Fig. \ref{fig:cno} shows the mean [N/C] ratios for giants in open clusters as a function of stellar mass. Prediction of stellar evolution models are shown for a phase either at the beginning or during core-He burning. This is thus after the first dredge-up has been completed and, for low-mass stars, after the extra mixing event that takes place at the luminosity bump. For intermediate-mass stars, this is also before any effect of the second dredge-up could become apparent. The predicted abundance ratios for the red giants are given with respect to the initial value. In this sense, they always depict the changes caused by the previous mixing events with respect to what the star had when it was formed.

For most clusters, the agreement of the [N/C] values with the expectation of the standard models \citep[][solid lines in Fig. \ref{fig:cno}]{2012MNRAS.427..127B,2012A&A...537A.146E,2012A&A...543A.108L} is very good. Only at masses above $\sim$ 5.0 M$_{\odot}$ a small deviation is seen. In this case, the [N/C] ratios determined in this work seem to be slightly below the model prediction (but still in agreement within the errors). At this stellar mass, the [N/C] ratios from the literature, on the other hand, seem to be slightly above the models. We note here that our oxygen abundances do not seem to deviate significantly from the solar value.

In Fig. \ref{fig:cno}, the error bars in the cluster values of [N/C] are the typical uncertainty ($\pm$ 0.14 dex) propagated from the errors in C and N listed in Table \ref{tab:cnosigma}. The cluster scatter was not computed as, for most clusters, good C and N abundances are available for one giant only (with exception of NGC 2099, where the [N/C] ratios in two giants agree to within 0.1 dex). One can still have an idea of the magnitude of a possible star-to-star scatter in [N/C] by looking at the points of different clusters with similar turn-off masses. From Fig. \ref{fig:cno}, we see that such variation is minimal. For example, for our four clusters with M$_{\rm TO}$ between 2.0 and 3.0 M$_{\odot}$, the mean [N/C] = +0.72 dex and the standard deviation is of $\pm$0.06 dex. This suggests that we did not detect a star-to-star scatter that needs rotation to be explained.

In intermediate-mass stars, the first dredge-up brings to the surface of the star CNO-cycle processed material. As a result, [C/Fe] decreases, [N/Fe] increases, and [O/Fe] remains practically identical with respect to its pre-dredge-up values. These stars do not have a luminosity-bump phase, as they quiescently start to burn helium at their core before the hydrogen-burning shell reaches the composition discontinuity left behind by the convective layer \citep[e.g.][]{2015MNRAS.453..666C}.

Interestingly, they go through a phenomenon not seen in low-mass stars, the so-called blue loops \citep[e.g.][]{2015MNRAS.447.2951W}. After He ignition in the core, the stars suffer a small decrease in its luminosity and heat up, moving to the left of the Hertzsprung-Russell diagram. During this loop, the stars go through a second and third crossing of the instability strip, pulsating as Cepheids. Chemical composition might help identify the first-crossing Cepheids, which are before the first dredge-up. Nevertheless, according to rotating stellar models, changes in the chemical composition might start even before the first dredge-up, making the identification of first-crossing Cepheids more complicated \citep{2014A&A...564A.100A}. For a recent discussion on first-crossing Cepheids see \citet{2016MNRAS.460.2077K}.

The literature [N/C] ratios for the Cepheids are shown in the right panel of Fig. \ref{fig:cno}. Stars with low [N/C] ratio are likely first-crossing Cepheids that have not completed the first dredge-up. This has already been reported by \citet{2011AJ....142...51L} in the case of SV Vul. This might be also the case for two stars in our sample, l Car and U Car. It is interesting to note that one of these stars (l Car) also has low [Na/Fe], probably indicating that the Na abundances was also not modified by the first dredge-up.

For the remaining Cepheids, most observed values concentrate around [N/C] $\sim$ +0.5 dex, slightly below the expectation of the models, where [N/C] is $\sim$ +0.6-0.7 dex for the standard case. A few higher values of [N/C] are also seen, for stars more massive than 6 M$_{\odot}$. RS Pup has the highest [N/C] in this sample ($\sim$ 1.1 dex). Such spread in [N/C] values has been seen in other works \citep[e.g.][]{2006A&A...449..655S}. The lack of non-LTE corrections could, in principle, create some scatter. However, we refer to Fig. 13 of \citet{2015MNRAS.446.3447L} which     shows that even with non-LTE C and N abundances the scatter remains. 

The stars with high [N/C] can in principle be explained by models including rotation-induced mixing with different efficiency. In the models of \citet{2012A&A...543A.108L}, the initial rotational velocity in the zero age main sequence (ZAMS) is of 144 and 156 km s$^{-1}$, for stars of 4 and 6 M$_{\odot}$, respectively \citep[see also][]{2014A&A...570C...2L}. In the models \citet{2012A&A...537A.146E}, the initial rotational velocity in the ZAMS is of 197 and 235 km s$^{-1}$, for stars of 4 and 7 M$_{\odot}$, respectively. The distinct initial velocities probably account for part of the difference seen in the rotating model curves of Fig. \ref{fig:cno}. The models of \citet{2012A&A...543A.108L} also included the effects of thermohaline mixing while the models of \citet{2012A&A...537A.146E} do not. Thermohaline mixing by itself has only a small impact in the [N/C] ratios at this metallicity, and only for stars below 2 M$_{\odot}$ \citep[see Fig. 19 of ][]{2010A&A...522A..10C}. Thermohaline mixing is not important for stars with masses above $\sim$ 2.0 M$_{\odot}$ before the second dredge-up.

The reasons behind the different mean level of the [N/C] ratio of the cluster giants to Cepheids is unclear. It might be connected to the differences in the methods of analysis and the features used to estimate the C and N abundances. While we used molecular features, atomic lines were used by \citet{2011AJ....142..136L}. A direct comparison is difficult because of the differences in temperature between Cepheids and our cluster giants. Molecular features are less prominent in the spectra of Cepheids and not commonly used to derive abundances, while the CNO atomic lines are weak or absent in the spectra of cool giants. Nevertheless, we remark that the conclusion that rotation-induced mixing is mostly not necessary to explain the [N/C] ratios is the same in the two completely independent samples, observed and analysed in completely independent ways.

\begin{figure*}
\centering
\includegraphics[height = 7cm]{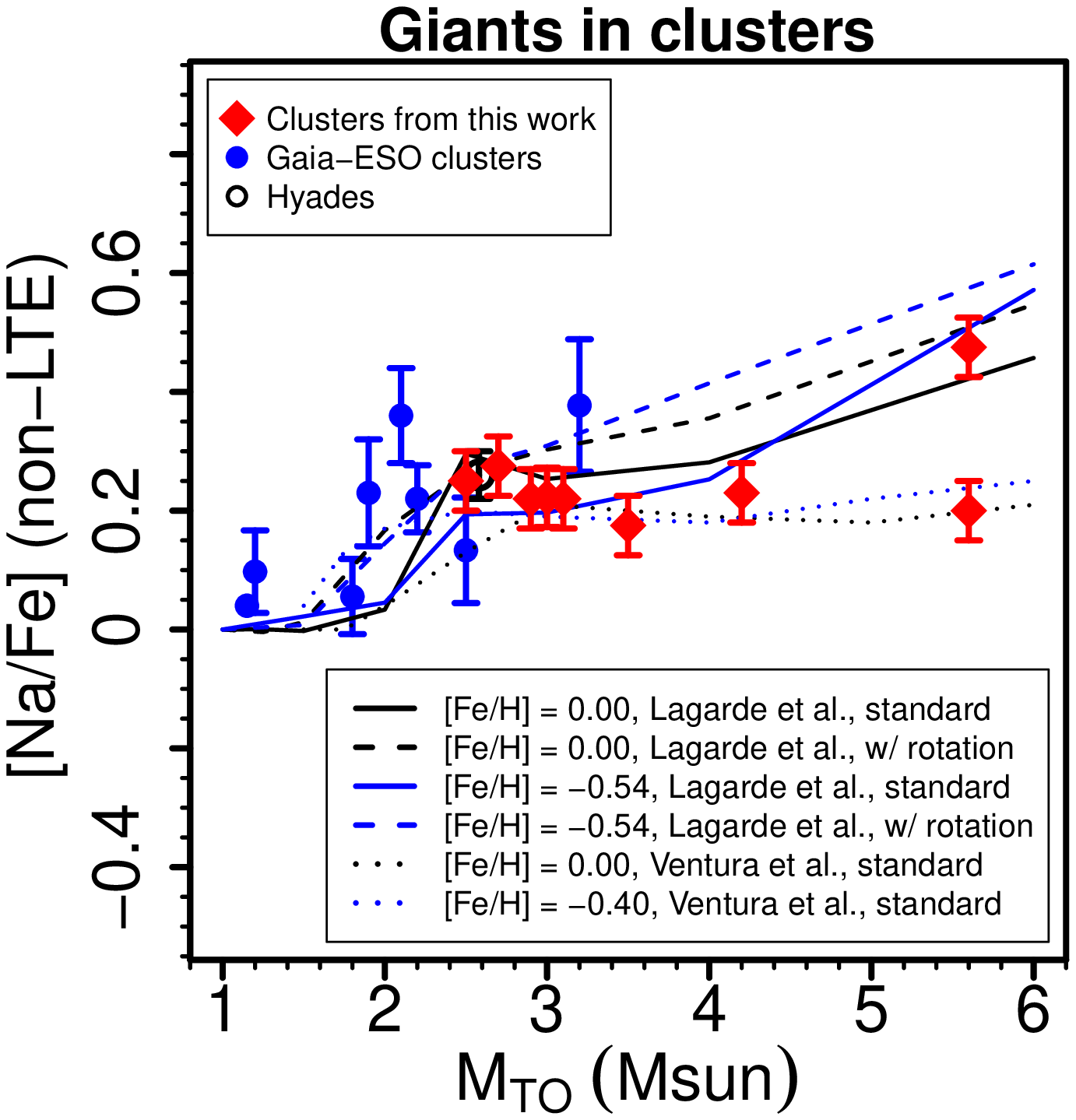}
\includegraphics[height = 7cm]{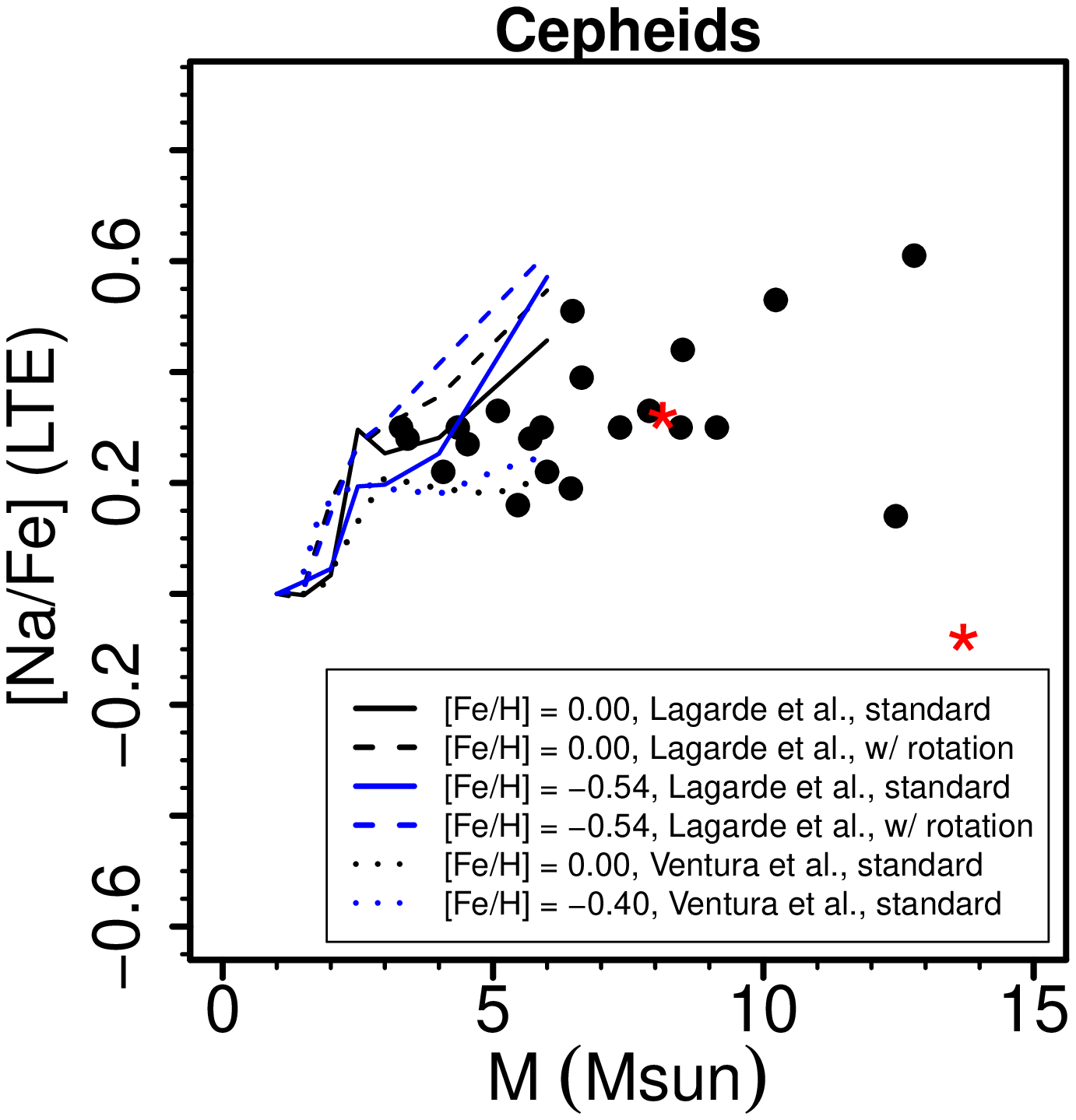}
 \caption{\emph{Left:} Mean cluster [Na/Fe] ratio, computed using only the member stars, as a function of stellar mass. The new observations are shown as red diamond symbols. The typical standard deviation of the mean [Na/Fe], $\pm$0.05 dex, is shown as an error bar. The data of the \emph{Gaia}-ESO clusters are from \citet{2016A&A...589A.115S}, \citet{2017A&A...598A..68O}, and \citet{2017A&A...601A..56T}. The data of the Hyades are from \citet{2012MNRAS.422.1562S}. We estimate the uncertainty in the turn-off masses to be less than $\pm$ 0.2 $M_{\odot}$. \emph{Right:} [Na/Fe] as a function of mass for the sample of selected Cepheids. Two Cepheids suspected to be at the first crossing are shown as red stars; the other Cepheids are indicated as solid circles. The abundances are in LTE. References are given in the text. The models extend only up to 6 M$_{\odot}$.}\label{fig:na}%
\end{figure*}
\begin{figure}
\centering
\includegraphics[height = 7cm]{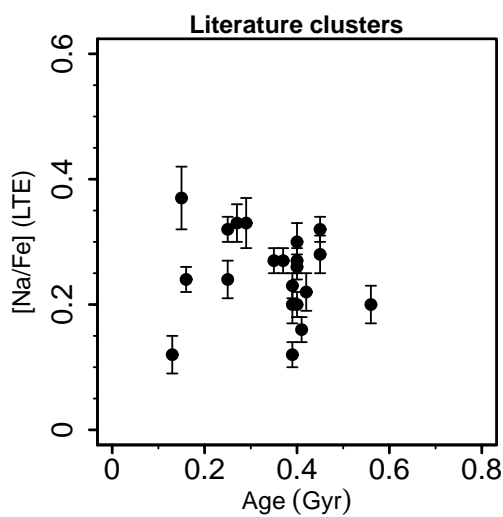}
 \caption{Literature results of [Na/Fe], in LTE, for open clusters as a function of cluster age. The mass range for ages between 100 Myr and 600 Myr is approximately between 5.5-2.5 M$_{\odot}$.}\label{fig:reddy}%
\end{figure}

\subsection{Sodium abundances}

The sodium abundances ([Na/Fe]) are shown as a function of stellar mass in Fig. \ref{fig:na}. Model curves from \citet{2012A&A...543A.108L} and \citet{2013MNRAS.431.3642V}\footnote{The model predictions are not available in the paper itself but were obtained directly from P. Ventura (2016, private communication).} are shown. As for the [N/C] ratios, the predictions are for a stage after the first dredge-up has been completed. We note that the largest differences between model predictions do not arise for the same set of models computed with or without rotation \citep{2012A&A...543A.108L}, but rather for different sets of models computed with different assumptions about the underlying stellar physics, \citet{2012A&A...543A.108L} versus\ \citet{2013MNRAS.431.3642V} . The models of  \citet{2012MNRAS.427..127B} and \citet{2012A&A...537A.146E} do not list the surface Na abundance.
 
The results for our open clusters are shown in the left panel. The giants below 2 M$_{\odot}$ are consistent with no Na enhancement \citep{2016A&A...589A.115S}. Between 2 and 3 M$_{\odot}$, the [Na/Fe] ratio increases. For the giants above 3 M$_{\odot}$ we can now establish that [Na/Fe] is constant around +0.2 dex for most cases. Thus, the new abundances that we determined tend to favour the standard models without rotation from \citet{2013MNRAS.431.3642V}.

The behaviour at $\sim$ 5.5 M$_{\odot}$ is again interesting. At this mass, the two standard models predict different [Na/Fe] surface abundances. While in the models of \citet{2012A&A...543A.108L} the surface [Na/Fe] is expected to start increasing with mass even without rotation, for the models of \citet{2013MNRAS.431.3642V} the trend with mass remains flat. The two clusters in our sample with turn-off masses of 5.6 M$_{\odot}$ behave differently. While for NGC 436, [Na/Fe] = +0.20 and for Trumpler 2, [Na/Fe] = +0.48. The Na abundance observed in NGC 436 only agrees with the models of \citet{2013MNRAS.431.3642V}. For Trumpler 2, the agreement is with the models of \citet{2012A&A...543A.108L}, even though we cannot decide among the models with or without rotation. 

\citet{2011MNRAS.417..649Z} determined [Na/Fe] = +0.45 (in LTE) for the same star that we analysed in Trumpler 2. We estimated a non-LTE correction in this case of $-$0.12 dex, adopting the atmospheric parameters and abundances determined by \citet{2011MNRAS.417..649Z}. Thus, our value is higher than that determined by \citet{2011MNRAS.417..649Z}. Nevertheless, the lower [Na/Fe] = +0.33 would still not bring Trumpler 2 into agreement with the \citet{2013MNRAS.431.3642V} model.

The [Na/Fe] ratios (in LTE) for the Cepheids are shown in the right panel of Fig. \ref{fig:na}. The mean value is [Na/Fe] = +0.30 dex. We estimated above that a typical non-LTE correction for Cepheids would be of the order of $-$0.05 dex. With this correction, the [Na/Fe] values of Cepheids would be brought to a closer agreement with the expectation of standard models. Basically, the mean trend with mass for Cepheids seems flat at [Na/Fe] $\sim$ +0.25 dex up to $\sim$9-10 M$_{\odot}$. This is in agreement with what is seen for the open cluster giants within the uncertainties. This suggests that for the majority of the Cepheids, rotation-induced mixing has not affected the post-dredge-up surface abundance of Na. A few cases of higher [Na/Fe] would likely require rotation-induced mixing to be explained.

Thus, the conclusion drawn for the Cepheids is the same as for the open cluster giants. Observationally, it seems that, in most cases, rotation-induced mixing does not affect the post dredge-up surface Na abundance significantly. Moreover, the models of \citet{2013MNRAS.431.3642V}, in which the trend of [Na/Fe] with stellar mass is flat, are preferred. Only a small number of stars seem to require rotation-induced mixing to be explained, even though in the standard models of \citet{2012A&A...543A.108L} the [Na/Fe] also increases with mass. Therefore, the conclusion drawn from the [N/C] ratios is supported by the Na abundances. The only discrepant case is Trumpler 2, where [Na/Fe] seems to require additional mixing that is not needed to explain the [N/C] ratio.

To further illustrate the trend of [Na/Fe] in clusters, we selected a few results from the literature coming from a single group \citep{2012MNRAS.419.1350R,2013MNRAS.431.3338R,2015MNRAS.450.4301R,2016MNRAS.463.4366R}. The LTE [Na/Fe] ratios are plotted against cluster age in Fig. \ref{fig:reddy}. The mass range should be equivalent to that of our new sample of clusters (between 2.5-5.5 M$_{\odot}$). In LTE, the average [Na/Fe] is +0.25 dex and no trend with age is apparent. A typical non-LTE correction of $-$0.10 dex, as derived for our sample, would bring the average to +0.15 dex (with range between +0.02 dex and +0.27 dex). The lack of trend and the mean value support the conclusion drawn from our own sample; strong effects of rotation-induced mixing are also not seen in this independent sample of cluster giants.

\subsection{Aluminium abundances}

\begin{figure}
\centering
\includegraphics[height = 7cm]{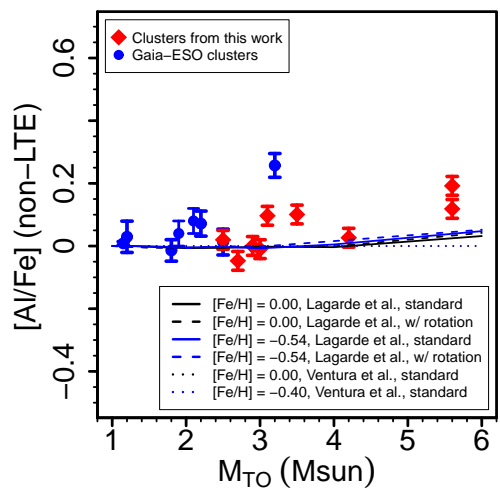}
 \caption{Mean [Al/Fe] cluster abundances, computed using only the member stars, as a function of stellar mass. The new observations are shown as red diamond symbols. The typical standard deviation of the mean [Al/Fe], $\pm$0.03 dex, is shown as an error bar. The data of the \emph{Gaia}-ESO clusters (blue symbols) are from \citet{2016A&A...589A.115S}, \citet{2017A&A...598A..68O}, and \citet{2017A&A...601A..56T}. The turn-off mass uncertainty is expected to be less than $\pm$ 0.2 $M_{\odot}$.}\label{fig:al}%
\end{figure}
\begin{table*}
 \caption[]{\label{tab:masscep} Comparison of stellar masses for the Cepheids that are also members of open clusters.}
\centering\small
\begin{tabular}{lccccccc}
\hline
\hline
Cepheid & Open cluster & [Fe/H] & Age & M$_{\rm TO}$ & M$_{\rm AGB}$ & M$_{\rm Cepheid}$ & Ref. \\
 & & & Myr & M$_{\odot}$ & M$_{\odot}$ & M$_{\odot}$ \\
\hline  
AQ Pup & \object{Turner 14} & -- & 32 & 8.6 & 8.9 & 9.14 &  (1) \\
BB Sgr    & \object{Collinder 394}         &  -- &  72   &     5.8  & 6.1 &  4.97 & (2) \\
CV Mon   & \object{VDB 1}   &  -- &        207   &     3.8  & 4.0 &  4.08 & (2) \\
EV Sct    & \object{NGC 6664}   &  -- &         79   &    5.6  & 5.8  & 5.46 & (2) \\
S Nor     & \object{NGC 6087}   & $-$0.01 &   89   &     5.2  & 5.5 &  6.64 & (2) \\
SU Cas &  \object{Alessi 95} & -- & 158 & 4.2 & 4.4 & 6.56 &  (3) \\
U Car    & \object{Feinstein 1} & -- & 16 & 12.8 & 13.5 & 8.15 &   (4) \\U Sgr     & \object{IC 4725}        & $-$0.26 &     93   &     5.2  & 5.5  & 4.50 & (2) \\
V340 Nor & \object{NGC 6067}  &  +0.14  &   93   &     5.2  & 5.5  & 5.09 & (2) \\
V Cen    & \object{NGC 5662}   & --  &      188   &     3.9  & 4.1 &  4.35 & (2) \\
VY Car  & \object{ASCC 61} & -- & 46 & 7.2 & 7.5 & 8.47 &  (4) \\
WZ Sgr   & \object{Turner 2}     & -- &         79   &     5.6  & 5.8 &  8.22 & (2) \\
\hline
\end{tabular}
\tablefoot{The parameters of the open clusters are from \citet{2013A&A...558A..53K} except for Alessi 95 \citep{2012MNRAS.422.2501T} and Turner 14 \citep{2012AJ....144..187T}. The stellar masses at the turn-off  and at the AGB are from PARSEC isochrones for the age of the open cluster and solar metallicity. The references that associate the Cepheids with an open cluster are (1) \citet{2012AJ....144..187T}; (2) \citet{2013MNRAS.434.2238A}; (3) \citet{2012MNRAS.422.2501T}; and (4) \citet{2015MNRAS.446.1268C}.}
\end{table*}

The aluminium abundances ([Al/Fe]) of the open cluster giants are shown as a function of stellar mass in Fig. \ref{fig:al}. The figure includes values for the \emph{Gaia}-ESO clusters reported in \citet{2016A&A...589A.115S}. The novelty in this plot is that all Al abundances have been corrected for non-LTE effects, which was not possible in \citet{2016A&A...589A.115S}. We did not collect abundances of Al for the Cepheids. 

Most [Al/Fe] ratios are distributed around values equal to zero, demonstrating that there was no change in surface Al abundance during the dredge-up. Two clusters between 3.0-3.5 M$_{\odot}$ have [Al/Fe] $\sim$ +0.10 dex. This small enhancement could still be caused by the errors of the analysis or perhaps comes from the cosmic scatter, as seen in the [Al/Fe] abundances of field stars in \citet{2016A&A...589A.115S}. The \emph{Gaia}-ESO cluster that stands out is NGC 6705. The comparison with the new data confirms the discussion in the Introduction that NGC 6705 seems to have anomalous abundances of Al \citep[see the discussion in][]{2015A&A...580A..85M,2016A&A...589A.115S}.

There is a suggestion of an upturn in [Al/Fe] for the masses around 5.5 M$_{\odot}$. As for the case of Na, it is Trumpler 2 the cluster with the highest [Al/Fe] ratio, +0.19 dex. If confirmed, this could indicate that at around this stellar mass, the dredge up reaches regions in which the temperature was high enough to activate the MgAl cycle. The models of \citet{2012A&A...543A.108L} themselves seem to hint that a small enhancement of Al becomes possible.

New observations of giants in this mass range are needed to test the upturn of the trend. Alternatively, observations of unevolved stars in Trumpler 2 could help to clarify whether the observed surface abundances are a result of deep mixing. For the moment, we can conclude that for stars below 5 M$_{\odot}$, the dredge up does not cause any change in the surface Al abundances.

\subsection{Cepheids in open clusters}

Twelve of the Cepheids in our selected sample are also members of open clusters. This offers the opportunity to compare the stellar masses obtained from the pulsation theory with masses obtained from isochrones. Such comparison might reveal whether inconsistencies among the two scales have any influence in the trends of surface abundances with mass. The list of cluster Cepheids and the values of stellar masses are given in Table \ref{tab:masscep}.

For seven stars, the agreement between the turn-off mass (which we adopt as the mass of cluster stars) and the pulsation mass is within 1 M$_{\odot}$, which we deem to be satisfactory. Moving these stars by less than 1 M$_{\odot}$ in the plots of Figs. \ref{fig:cno} and \ref{fig:na} has no effect on the trends of abundance with mass.

For five stars (S Nor, SU Cas, U Car, VY Car, and WZ Sgr) there is a difference of more than 1 M$_{\odot}$ between masses. In all cases, except U Car, the pulsation mass is bigger than the turn-off mass. Nevertheless, even decreasing the masses of these stars by $\sim$ 2.5 M$_{\odot}$ would still not change the general behaviour of the abundances as a function of mass. Finally, increasing the mass of U Car from $\sim$ 8 to $\sim$ 13 M$_{\odot}$ would, if anything, help stress the disagreement with the models including rotation. This star has [Na/Fe] = +0.32 dex and [N/C] = +0.59 dex, which are below what seems to be expected in the case of models including rotation-induced mixing.

\subsection{Rotation history of the giants and Cepheids}

Though the observed abundances seem to favour non-rotating models, the main-sequence progenitors of the observed giants and Cepheids should actually be
fast-rotating stars. With a mass range between 2.5 M$_{\odot}$ to 10 M$_{\odot}$, the giants and Cepheids of the sample were main-sequence stars of spectral types B or early A. These early-type main-sequence stars are know to be fast rotators \citep[][]{1949ApJ...110..498S,2002ApJ...573..359A,2010ApJ...722..605H}.

A quantitative example is given by the analysis of stars in the young open cluster NGC 6067 (with age $\sim$ 50-150 Myr) by \citet[][]{2017MNRAS.469.1330A}. \citet{2017MNRAS.469.1330A} found 25 B-type turn-off star cluster members to have values of $v~\sin~i$ between 0.40 and 281 km s$^{-1}$ with mean $\sim$ 135 $\pm$ 88 (s.d.) km s$^{-1}$. As we mentioned above, the various models with rotation shown in Figs. \ref{fig:cno} and \ref{fig:na} assume initial rotational velocities in the ZAMS between 144 and 235 km s$^{-1}$ for stars between 4 and 7 M$_{\odot}$ \citep{2012A&A...543A.108L,2014A&A...570C...2L,2012A&A...537A.146E}.

Ignoring the angle of inclination, the above $v~\sin~i$ distribution would suggest that the surface abundances predicted by the rotating models of \citet{2012A&A...543A.108L} should be the norm. Giants with abundances in agreement with standard models should be a minority and, additionally, some giants should have undergone mixing as strong as predicted by the models of \citet{2012A&A...537A.146E}. The angle of inclination can make fast rotators look like slow rotators, but never the opposite. So the distribution of $v~\sin~i$ is a distribution of lower limits of the true rotational velocity. Thus, the models with rotation should actually be the minimum expected to be seen in the observations. That, however, is not the case in the samples discussed in this work.

In any case, there is evidence of intermediate-mass giants with large [N/C] ratios that can be explained only taking into account rotation-induced mixing \citep[e.g.][]{2006A&A...449..655S,2015MNRAS.446.3447L}. This likely includes the weak G-band stars \citep{1973AJ.....78..687B}, which seem to be intermediate-mass giants showing signs of very strong mixing \citep{2012A&A...538A..68P,2016A&A...587A..42P,2013ApJ...765..155A}. Our sample of Cepheids itself contains a few stars with high values of [N/C] and [Na/Fe] that might require rotation-induced mixing.

The fact that no strong signs of rotation-induced mixing are seen could be explained if our stellar samples are made preferentially of objects that evolved from slow rotators. If fast-rotating main-sequence stars give origin to fast-rotating giants or Cepheids, they could have been discarded from the analysis (as we did with star Cr 421 566 in our sample). We might test this looking at the distribution of $v~\sin~i$ in a large sample of bright giants to understand whether there is a significant number of fast-rotating stars that would be discarded from abundance analyses.

The catalogue of \citet{2014A&A...561A.126D} lists $v~\sin~i$ values for 100 single-lined field G- or K-type supergiants (luminosity class Ib, II or II-III). The mean $v~\sin~i$ = 3.8 $\pm$ 3.0 (s.d.) km s$^{-1}$ with a range between 1.0 km s$^{-1}$ and 11.8 km s$^{-1}$. \citet{2006A&A...453..309N} found $v~\sin~i$ $<$ 16 km s$^{-1}$ in nine Cepheids and \citet{2013PhDT.......363A} found an average equatorial velocity of 12.3 km s$^{-1}$ in 97 Cepheids. These numbers suggest that very fast rotators are actually rare among intermediate-mass giants and Cepheids. Based on this, it does seem that fast-rotating main-sequence stars slow down and become slow-rotating giants. Therefore, a bias towards selecting stars that evolved from slow-rotating main-sequence objects in our samples is unlikely. 

An additional possibility is that the models are overestimating the effects of rotation-induced mixing. Indeed, such possibility was already raised by \citet{2014A&A...564A.100A}, in a comparison between models and CNO abundances of Cepheids, and by \citet{2013MNRAS.433.1246S}, who looked at CNO abundances in blue supergiants. Our results support their conclusion. We note here that the Li abundance derived for star Cr 421 566 (Section \ref{sec:lirich}) is also consistent with the same idea. This giant, with $\sim$3.5 M$_{\odot,}$ has A(Li) $\sim$ 1.1 dex. The models without rotation by \citet{2012A&A...543A.108L} indicated A(Li) $\sim$ 1.3 dex for stars of both 3.0 and 4.0 M$_{\odot}$, after the first dredge-up. In the case including rotation, the predicted surface abundances are A(Li) = 0.2 and 0.0 dex for giants of 3.0 and 4.0 M$_{\odot}$, respectively. This predicted depletion is much stronger than what is observed in star Cr 421 566.

Based on the samples analysed here, it seems that effects of rotation-induced mixing on the surface abundances of C, N, and Na are mostly overestimated by the evolutionary models of intermediate-mass giants and supergiants. Such a conclusion comes from the average behaviour of the abundances in two different samples of stars that were analysed independently of each other.

%-------------------------------------------------------------------

\section{Summary}\label{sec:end}

We obtained new high-resolution high S/N ratio optical spectra of 20 red giants in the field of 10 open clusters with ages between 80 and 610 Myr (equivalent to a mass range between 2.5 M$_{\odot}$ and 5.6 M$_{\odot}$). We derived chemical abundances of 22 elements using EWs and of C, N, and O using spectrum synthesis. We found that 17 of the 20 giants are likely members of nine clusters.

Our main aim was to study the surface abundances of Na and Al and investigate the trends with stellar mass to understand how they are affected by deep mixing during the first dredge-up. To complement the sample, we selected 32 Cepheids from the literature with accurate values of masses and abundances. This extends the probed mass range to 14 M$_{\odot}$.

We found that rotation-induced mixing seems to play a small role in the changes in surface abundances during the first dredge-up. In the majority of the stars, both the [N/C] ratios and Na abundances do not show signs of being affected by rotation-induced mixing. Indeed, the observed [N/C] ratios of Cepheids even seem to be below the expectation of standard non-rotating models, although we cannot exclude the presence of biases in the comparison. We found no significant change in Al abundances for giants below 5 M$_{\odot}$. There is perhaps a hint of small change when stellar mass reaches close to 6 M$_{\odot}$. This needs further analysis to be confirmed. The agreement of [Na/Fe] surface abundances with standard models supports the findings of \citet{1994PASJ...46..395T}, \citet{1995ApJ...451..298E}, \citet{2005PASP..117.1173K}, and \citet{2013MNRAS.432..769T}.

It seems surprising that signs of rotation-induced mixing are not detected, given that the main-sequence progenitors of these giants were B- or early A-type stars which are known to be fast rotators. Additionally, it seems that rotation is needed to explain other properties of Cepheids \citep[e.g.][]{2014A&A...564A.100A,2016A&A...591A...8A}.

Although models with rotation-induced mixing seem to, on average, overestimate the effects of rotation in the surface abundances of the samples discussed in this work, we remark that there are observations of [N/C] in intermediate-mass giants that require rotation to be explained \citep[e.g.][]{2006A&A...449..655S,2012A&A...538A..68P,2016A&A...587A..42P,2013ApJ...765..155A,2015MNRAS.446.3447L}. Enlarging the sample of intermediate-mass stars with accurate abundances and masses is needed to further investigate the discrepancy with respect to rotating evolutionary models.

\begin{acknowledgements}
RS acknowledges support from the Polish Ministry of Science and Higher Education. AB thanks ARI Heidelberg, where part of this project was discussed, for the hospitality. BL acknowledges  support by Sonderforschungsbereich SFB 881 "The Milky Way System" (subproject A5) of the German Research Foundation (DFG). The authors thank Thomas Nordlander for providing the IDL routine that interpolates the non-LTE corrections for the Al abundances. Based on observations made with the Nordic Optical Telescope, operated by the Nordic Optical Telescope Scientific Association at the Observatorio del Roque de los Muchachos, La Palma, Spain, of the Instituto de Astrofisica de Canarias. This research has made use of NASA's Astrophysics Data System; the SIMBAD database, operated at CDS, Strasbourg, France; and the VizieR catalogue access tool, CDS, Strasbourg, France. The original description of the VizieR service was published in \citet{2000A&AS..143...23O}; the WEBDA database, operated at the Department of Theoretical Physics and Astrophysics of the Masaryk University. The analysis has made use of {\sf R} \citep{Rcore} and the {\sf R} package {\sf gplots} \citep{gplots}.

\end{acknowledgements}

\bibliographystyle{aa} 
\bibliography{../../smiljanic}

\begin{appendix}

\section{Chemical abundances of the red giants in open clusters}

%
%\begin{table*}
% \caption[]{\label{tab:abun} Abundances based on equivalent widths.}
%\centering
%\begin{tabular}{lccccccccccccccccccc}
%\hline
%\hline
\begin{sidewaystable*}
\caption{Chemical abundances based on equivalent widths for the giants in open clusters. The Na and Al abundances are listed in non-LTE.}\label{tab:abun}
\centering\small
\begin{tabular}{lccccccc}
\hline\hline             
 {[Elem/Fe]} & Cr 421 466 & Cr 421 529 & Cr 421 566 & NGC 436 482 & NGC 1342 4 & NGC 1342 7 & NGC 1528 1009  \\ 
\hline
{[FeI/H]} & $-$0.12 $\pm$ 0.12 (197) & $-$0.07 $\pm$ 0.10 (210) & $-$0.03 $\pm$ 0.15 (148) & $-$0.15 $\pm$ 0.12 (163) & $-$0.05 $\pm$ 0.11 (277) & $-$0.10 $\pm$ 0.13 (303) & $-$0.05 $\pm$ 0.13 (257) \\
{[FeII/H]} & $-$0.22 $\pm$ 0.10 (20) & $-$0.19 $\pm$ 0.10 (21) & $-$0.18 $\pm$ 0.11 (14) & $-$0.24 $\pm$ 0.18 (20) & $-$0.20 $\pm$ 0.10 (23) & $-$0.26 $\pm$ 0.09 (24) & $-$0.14 $\pm$ 0.14 (23) \\
{[AlI/Fe]} &  0.05 $\pm$ 0.01 (2) &  0.02 $\pm$ 0.05 (2) &  0.23 $\pm$ 0.00 (2) &  0.12 $\pm$ 0.04 (2) &  $-$0.01 $\pm$ 0.02 (2) &  $-$0.01 $\pm$ 0.03 (2) &  0.10 $\pm$ 0.02 (2) \\
{[BaII/Fe]} &    -- &    -- &    -- &    -- &    -- &    --  &  --  \\
{[CaI/Fe]} & $-$0.01 $\pm$ 0.07 (5) & $-$0.06 $\pm$ 0.06 (4) &  0.00 $\pm$ 0.07 (5) &  0.09 $\pm$ 0.08 (3) & $-$0.02 $\pm$ 0.06 (12) &  0.02 $\pm$ 0.10 (18) &  0.01 $\pm$ 0.08 (6) \\
{[CeII/Fe]} & $-$0.05 $\pm$ 0.11 (5) & $-$0.01 $\pm$ 0.12 (6) & $-$0.13 $\pm$ 0.04 (2) &  0.12 $\pm$ 0.08 (4) &  0.00 $\pm$ 0.05 (4) & $-$0.12 $\pm$ 0.07 (3) &  0.05 $\pm$ 0.06 (4) \\
{[CoI/Fe]} & $-$0.01 $\pm$ 0.13 (12) &  0.01 $\pm$ 0.11 (13) & $-$0.04 $\pm$ 0.07 (7) &  0.01 $\pm$ 0.06 (10) & $-$0.03 $\pm$ 0.10 (14) & $-$0.07 $\pm$ 0.09 (15) &  0.07 $\pm$ 0.13 (14) \\
{[CrI/Fe]} &  0.06 $\pm$ 0.12 (15) &  0.00 $\pm$ 0.09 (15) &  0.11 $\pm$ 0.15 (4) &  0.10 $\pm$ 0.12 (15) & $-$0.06 $\pm$ 0.10 (12) & $-$0.02 $\pm$ 0.10 (12) &  0.02 $\pm$ 0.09 (15) \\
{[CrII/Fe]} & $-$0.13 $\pm$ 0.00 (1) & $-$0.09 $\pm$ 0.00 (1) &  0.72 $\pm$ 0.00 (1) &  0.16 $\pm$ 0.00 (1) & $-$0.03 $\pm$ 0.01 (2) &    -- & $-$0.08 $\pm$ 0.00 (1) \\
{[CuI/Fe]} & $-$0.30 $\pm$ 0.00 (1) & $-$0.11 $\pm$ 0.33 (2) &  0.33 $\pm$ 1.06 (2) & $-$0.25 $\pm$ 0.00 (1) & $-$0.38 $\pm$ 0.06 (2) & $-$0.65 $\pm$ 0.12 (2) &  0.01 $\pm$ 0.28 (2) \\
{[EuII/Fe]} &  0.02 $\pm$ 0.10 (2) &  0.01 $\pm$ 0.10 (2) &  0.07 $\pm$ 0.28 (2) &  0.11 $\pm$ 0.04 (2) & $-$0.01 $\pm$ 0.24 (2) &  0.01 $\pm$ 0.20 (2) &  0.07 $\pm$ 0.00 (1) \\
{[LaII/Fe]} & $-$0.07 $\pm$ 0.06 (3) & $-$0.09 $\pm$ 0.02 (3) & $-$0.32 $\pm$ 0.24 (2) & $-$0.03 $\pm$ 0.07 (4) & $-$0.06 $\pm$ 0.09 (4) & $-$0.17 $\pm$ 0.04 (3) & $-$0.12 $\pm$ 0.09 (4) \\
{[MgI/Fe]} &  0.14 $\pm$ 0.06 (2) &  0.20 $\pm$ 0.00 (2) &  0.39 $\pm$ 0.00 (1) &  0.18 $\pm$ 0.01 (2) &  0.06 $\pm$ 0.04 (2) &  0.13 $\pm$ 0.01 (2) &  0.12 $\pm$ 0.04 (2) \\
{[NaI/Fe]} &  0.15 $\pm$ 0.00 (2) &  0.19 $\pm$ 0.05 (2) &  0.19 $\pm$ 0.10 (2) &  0.20 $\pm$ 0.01 (2) &  0.28 $\pm$ 0.01 (2) &  0.17 $\pm$ 0.04 (3) &  0.22 $\pm$ 0.02 (3) \\
{[NiI/Fe]} & $-$0.01 $\pm$ 0.10 (13) & $-$0.06 $\pm$ 0.06 (15) & $-$0.09 $\pm$ 0.14 (13) & $-$0.04 $\pm$ 0.08 (10) & $-$0.07 $\pm$ 0.09 (23) & $-$0.08 $\pm$ 0.08 (24) & $-$0.04 $\pm$ 0.10 (18) \\
{[ScI/Fe]} & $-$0.10 $\pm$ 0.01 (4) & $-$0.08 $\pm$ 0.01 (3) &  0.10 $\pm$ 0.00 (1) & $-$0.07 $\pm$ 0.13 (3) & $-$0.11 $\pm$ 0.03 (3) & $-$0.12 $\pm$ 0.01 (2) & $-$0.10 $\pm$ 0.06 (4) \\
{[ScII/Fe]} &  0.06 $\pm$ 0.08 (4) &  0.03 $\pm$ 0.10 (5) & $-$0.27 $\pm$ 0.48 (2) &  0.05 $\pm$ 0.14 (5) & $-$0.05 $\pm$ 0.11 (5) & $-$0.03 $\pm$ 0.11 (7) &  0.04 $\pm$ 0.08 (7) \\
{[SI/Fe]} &    -- &    -- & $-$0.10 $\pm$ 0.46 (2) &    -- &    -- &    -- &    -- \\
{[SiI/Fe]} &  0.06 $\pm$ 0.09 (10) &  0.01 $\pm$ 0.08 (9) &  0.04 $\pm$ 0.12 (7) &  0.08 $\pm$ 0.10 (7) & $-$0.05 $\pm$ 0.05 (10) & $-$0.07 $\pm$ 0.06 (10) &  0.09 $\pm$ 0.08 (9) \\
{[SiII/Fe]} &  0.04 $\pm$ 0.20 (2) & $-$0.22 $\pm$ 0.00 (1) &  0.07 $\pm$ 0.10 (2) &  0.54 $\pm$ 0.00 (1) & $-$0.38 $\pm$ 0.07 (2) & $-$0.18 $\pm$ 0.11 (2) &  0.04 $\pm$ 0.30 (2) \\
{[SrI/Fe]} &  0.74 $\pm$ 0.16 (2) &  0.83 $\pm$ 0.16 (2) &    -- &  0.61 $\pm$ 0.01 (2) &    -- &    -- &  0.69 $\pm$ 0.27 (2) \\
{[TiI/Fe]} &  0.02 $\pm$ 0.06 (38) & $-$0.01 $\pm$ 0.06 (35) &  0.07 $\pm$ 0.08 (18) &  0.09 $\pm$ 0.10 (29) &  0.03 $\pm$ 0.08 (43) &  0.02 $\pm$ 0.08 (41) &  0.03 $\pm$ 0.07 (39) \\
{[TiII/Fe]} &  0.03 $\pm$ 0.14 (6) & $-$0.04 $\pm$ 0.08 (7) & $-$0.26 $\pm$ 0.04 (2) &  0.07 $\pm$ 0.21 (6) & $-$0.04 $\pm$ 0.09 (7) &  0.04 $\pm$ 0.12 (10) &  0.03 $\pm$ 0.11 (8) \\
{[VI/Fe]} & $-$0.04 $\pm$ 0.05 (13) & $-$0.05 $\pm$ 0.05 (16) &  0.01 $\pm$ 0.11 (12) &  0.13 $\pm$ 0.04 (8) & $-$0.06 $\pm$ 0.06 (19) & $-$0.02 $\pm$ 0.13 (20) &  0.06 $\pm$ 0.09 (13) \\
{[YII/Fe]} & $-$0.05 $\pm$ 0.08 (8) & $-$0.11 $\pm$ 0.10 (8) & $-$0.31 $\pm$ 0.16 (4) & $-$0.01 $\pm$ 0.11 (6) & $-$0.12 $\pm$ 0.11 (11) & $-$0.06 $\pm$ 0.22 (11) & $-$0.16 $\pm$ 0.09 (8) \\
{[ZnI/Fe]} &  0.07 $\pm$ 0.78 (2) & $-$0.44 $\pm$ 0.00 (1) & $-$0.55 $\pm$ 0.00 (1) & $-$0.49 $\pm$ 0.00 (1) & $-$0.14 $\pm$ 0.36 (2) & $-$0.09 $\pm$ 0.27 (2) & $-$0.34 $\pm$ 0.00 (1) \\
{[ZrI/Fe]} &  0.10 $\pm$ 0.06 (7) &  0.13 $\pm$ 0.01 (4) &    -- &  0.19 $\pm$ 0.03 (7) &    -- &    -- &  0.10 $\pm$ 0.05 (6) \\
{[ZrII/Fe]} &  0.43 $\pm$ 0.11 (2) &  0.42 $\pm$ 0.12 (2) &  0.28 $\pm$ 0.75 (2) &  0.53 $\pm$ 0.01 (2) &  0.39 $\pm$ 0.13 (2) &  0.42 $\pm$ 0.00 (1) &  0.42 $\pm$ 0.15 (2) \\
\hline
\end{tabular}
\end{sidewaystable*}

\setcounter{table}{0}

\begin{sidewaystable*}
\caption{Cont.: Chemical abundances based on EWs for the giants in open clusters. The Na abundances are listed in non-LTE.}%\label{tab:abun}
\centering\small
\begin{tabular}{lccccccc}
\hline\hline             
 {[Elem/Fe]} & NGC 1528 4876 & NGC 1662 1 & NGC 1662 2 & NGC 2099 1898 & NGC 2099 2520 & NGC 2281 55 & NGC 2281 63 \\
\hline
{[FeI/H]} & $-$0.47 $\pm$ 0.14 (256) & $-$0.01 $\pm$ 0.12 (270) & $-$0.03 $\pm$ 0.13 (325) &  0.04 $\pm$ 0.12 (251) &  0.10 $\pm$ 0.13 (242) &  0.00 $\pm$ 0.11 (301) & $-$0.07 $\pm$ 0.11 (198) \\
{[FeII/H]} & $-$0.62 $\pm$ 0.14 (14) & $-$0.17 $\pm$ 0.10 (22) & $-$0.17 $\pm$ 0.09 (24) & $-$0.11 $\pm$ 0.10 (22) & $-$0.06 $\pm$ 0.10 (23) & $-$0.16 $\pm$ 0.10 (28) & $-$0.16 $\pm$ 0.12 (20) \\
{[AlI/Fe]} &  0.23 $\pm$ 0.05 (2) &  $-$0.03 $\pm$ 0.09 (2) & $-$0.07 $\pm$ 0.01 (2) &  $-$0.02 $\pm$ 0.01 (2) &  0.02 $\pm$ 0.06 (2) &  $-$0.03 $\pm$ 0.03 (2) &  0.07 $\pm$ 0.01 (2) \\
{[BaII/Fe]} &    --  &    --  & $-$0.94 $\pm$ 0.00 (1) &    --  &    --  & $-$0.49 $\pm$ 0.63 (2) &    --  \\
{[CaI/Fe]} &  0.25 $\pm$ 0.11 (3) & $-$0.01 $\pm$ 0.06 (12) & $-$0.06 $\pm$ 0.09 (16) &  0.00 $\pm$ 0.06 (9) & $-$0.04 $\pm$ 0.08 (9) & $-$0.01 $\pm$ 0.04 (13) & $-$0.03 $\pm$ 0.11 (3) \\
{[CeII/Fe]} & $-$0.08 $\pm$ 0.12 (3) & $-$0.02 $\pm$ 0.12 (4) & $-$0.08 $\pm$ 0.08 (3) & $-$0.01 $\pm$ 0.09 (3) & $-$0.08 $\pm$ 0.12 (4) & $-$0.13 $\pm$ 0.07 (3) &  0.05 $\pm$ 0.08 (5) \\
{[CoI/Fe]} &  0.17 $\pm$ 0.15 (13) & $-$0.01 $\pm$ 0.09 (14) & $-$0.05 $\pm$ 0.10 (14) &  0.00 $\pm$ 0.09 (14) &  0.02 $\pm$ 0.12 (16) & $-$0.04 $\pm$ 0.09 (13) &  0.01 $\pm$ 0.08 (14) \\
{[CrI/Fe]} &  0.11 $\pm$ 0.10 (18) & $-$0.04 $\pm$ 0.13 (13) & $-$0.03 $\pm$ 0.11 (11) &  0.01 $\pm$ 0.09 (13) & $-$0.02 $\pm$ 0.14 (13) & $-$0.04 $\pm$ 0.05 (16) &  0.04 $\pm$ 0.10 (16) \\
{[CrII/Fe]} &  0.04 $\pm$ 0.00 (1) &    --  &  0.01 $\pm$ 0.06 (2) &    --  &  0.11 $\pm$ 0.04 (2) & $-$0.09 $\pm$ 0.02 (2) & $-$0.08 $\pm$ 0.00 (1) \\
{[CuI/Fe]} &  0.06 $\pm$ 0.00 (1) & $-$0.43 $\pm$ 0.07 (2) & $-$0.41 $\pm$ 0.10 (2) & $-$0.37 $\pm$ 0.03 (2) & $-$0.33 $\pm$ 0.05 (2) & $-$0.40 $\pm$ 0.10 (2) & $-$0.26 $\pm$ 0.00 (1) \\
{[EuII/Fe]} & $-$0.09 $\pm$ 0.06 (2) &  0.03 $\pm$ 0.18 (2) & $-$0.03 $\pm$ 0.22 (2) & $-$0.06 $\pm$ 0.08 (2) &  0.00 $\pm$ 0.10 (2) & $-$0.05 $\pm$ 0.32 (2) &  0.00 $\pm$ 0.06 (2) \\
{[LaII/Fe]} & $-$0.27 $\pm$ 0.04 (3) & $-$0.12 $\pm$ 0.04 (3) & $-$0.05 $\pm$ 0.04 (3) & $-$0.04 $\pm$ 0.15 (4) & $-$0.07 $\pm$ 0.25 (4) & $-$0.26 $\pm$ 0.05 (4) & $-$0.05 $\pm$ 0.01 (3) \\
{[MgI/Fe]} &  0.20 $\pm$ 0.01 (2) &  0.07 $\pm$ 0.03 (2) &  0.14 $\pm$ 0.02 (2) &  0.21 $\pm$ 0.10 (2) &  0.16 $\pm$ 0.06 (2) &  0.09 $\pm$ 0.04 (3) &  0.11 $\pm$ 0.02 (2) \\
{[NaI/Fe]} &  0.25 $\pm$ 0.02 (3) &  0.27 $\pm$ 0.02 (2) & 0.28 $\pm$ 0.03 (2) &  0.21 $\pm$ 0.03 (2) &  0.23 $\pm$ 0.01 (2) &  0.19 $\pm$ 0.05 (4) &  0.30 $\pm$ 0.05 (2) \\
{[NiI/Fe]} &  0.00 $\pm$ 0.10 (19) & $-$0.04 $\pm$ 0.08 (21) & $-$0.08 $\pm$ 0.08 (19) & $-$0.03 $\pm$ 0.09 (19) & $-$0.02 $\pm$ 0.10 (20) & $-$0.08 $\pm$ 0.07 (23) & $-$0.04 $\pm$ 0.08 (13) \\
{[ScI/Fe]} &  0.18 $\pm$ 0.11 (4) & $-$0.11 $\pm$ 0.00 (3) & $-$0.18 $\pm$ 0.01 (2) & $-$0.11 $\pm$ 0.02 (4) & $-$0.12 $\pm$ 0.03 (3) & $-$0.13 $\pm$ 0.01 (2) & $-$0.08 $\pm$ 0.10 (5) \\
{[ScII/Fe]} &  0.15 $\pm$ 0.08 (6) &  0.00 $\pm$ 0.10 (6) &  0.03 $\pm$ 0.08 (7) &  0.03 $\pm$ 0.09 (6) &  0.07 $\pm$ 0.03 (5) & $-$0.03 $\pm$ 0.08 (9) &  0.09 $\pm$ 0.10 (5) \\
{[SI/Fe]} &    --  &    --  &    --  & $-$0.18 $\pm$ 0.00 (1) &    --  &    --  &    --  \\
{[SiI/Fe]} &  0.04 $\pm$ 0.10 (10) & $-$0.08 $\pm$ 0.08 (11) & $-$0.06 $\pm$ 0.08 (12) &  0.01 $\pm$ 0.06 (9) & $-$0.05 $\pm$ 0.10 (11) & $-$0.06 $\pm$ 0.05 (11) &  0.07 $\pm$ 0.10 (10) \\
{[SiII/Fe]} &    --  & $-$0.26 $\pm$ 0.02 (2) & $-$0.18 $\pm$ 0.10 (2) & $-$0.13 $\pm$ 0.16 (2) & $-$0.19 $\pm$ 0.00 (1) & $-$0.21 $\pm$ 0.12 (2) &    --  \\
{[SrI/Fe]} &  0.65 $\pm$ 0.09 (2) &  0.63 $\pm$ 0.00 (1) &  0.66 $\pm$ 0.00 (1) &  0.44 $\pm$ 0.00 (1) &    --  &    --  &  0.62 $\pm$ 0.06 (2) \\{[TiI/Fe]} &  0.28 $\pm$ 0.10 (27) &  0.02 $\pm$ 0.09 (40) &  0.03 $\pm$ 0.07 (43) &  0.01 $\pm$ 0.08 (45) &  0.01 $\pm$ 0.09 (41) & $-$0.02 $\pm$ 0.07 (56) &  0.03 $\pm$ 0.07 (41) \\
{[TiII/Fe]} &  0.12 $\pm$ 0.20 (11) &  0.07 $\pm$ 0.14 (9) &  0.07 $\pm$ 0.16 (10) & $-$0.03 $\pm$ 0.11 (9) &  0.02 $\pm$ 0.12 (8) & $-$0.03 $\pm$ 0.08 (10) &  0.02 $\pm$ 0.10 (8) \\
{[VI/Fe]} &  0.30 $\pm$ 0.08 (9) & $-$0.03 $\pm$ 0.10 (18) & $-$0.05 $\pm$ 0.08 (18) & $-$0.01 $\pm$ 0.09 (21) & $-$0.01 $\pm$ 0.09 (21) & $-$0.10 $\pm$ 0.06 (20) &  0.07 $\pm$ 0.04 (10) \\
{[YII/Fe]} & $-$0.07 $\pm$ 0.23 (9) & $-$0.09 $\pm$ 0.14 (11) & $-$0.05 $\pm$ 0.15 (9) & $-$0.17 $\pm$ 0.13 (9) & $-$0.08 $\pm$ 0.19 (10) & $-$0.09 $\pm$ 0.10 (10) & $-$0.08 $\pm$ 0.07 (6) \\
{[ZnI/Fe]} & $-$0.57 $\pm$ 0.00 (1) & $-$0.08 $\pm$ 0.42 (2) & $-$0.09 $\pm$ 0.32 (2) &  0.04 $\pm$ 0.56 (2) & $-$0.50 $\pm$ 0.00 (1) & $-$0.40 $\pm$ 0.00 (1) & $-$0.46 $\pm$ 0.00 (1) \\
{[ZrI/Fe]} &  0.13 $\pm$ 0.07 (9) &    --  &    --  &  0.14 $\pm$ 0.03 (2) &  0.11 $\pm$ 0.00 (1) & $-$0.03 $\pm$ 0.00 (1) &  0.12 $\pm$ 0.04 (8) \\
{[ZrII/Fe]} &  0.50 $\pm$ 0.11 (2) &  0.02 $\pm$ 0.00 (1) &  0.16 $\pm$ 0.10 (2) &  0.39 $\pm$ 0.15 (2) &    --  &  0.32 $\pm$ 0.11 (2) &  0.52 $\pm$ 0.04 (2) \\
\hline
\end{tabular}
\end{sidewaystable*}

\setcounter{table}{0}

\begin{sidewaystable*}
\caption{Cont.: Chemical abundances based on EWs for the giants in open clusters. The Na abundances are listed in non-LTE.}%\label{tab:abun}
\centering\small
\begin{tabular}{lccccc}
\hline\hline             
 {[Elem/Fe]} & NGC 2281 74 & NGC6709 208 & NGC6709 1998 & St2 0160 & Tr2 01 \\
\hline
{[FeI/H]} &  0.00 $\pm$ 0.11 (270) & $-$0.01 $\pm$ 0.12 (199) & $-$0.04 $\pm$ 0.11 (194) & $-$0.05 $\pm$ 0.11 (308) & $-$0.24 $\pm$ 0.13 (169) \\
{[FeII/H]} & $-$0.16 $\pm$ 0.10 (16) & $-$0.10 $\pm$ 0.16 (21) & $-$0.17 $\pm$ 0.10 (15) & $-$0.21 $\pm$ 0.12 (28) & $-$0.32 $\pm$ 0.19 (20) \\
{[AlI/Fe]} &  0.07 $\pm$ 0.01 (2) &  0.03 $\pm$ 0.01 (2) &  0.06 $\pm$ 0.04 (2) &  $-$0.03 $\pm$ 0.04 (2) &  0.19 $\pm$ 0.03 (2) \\
{[BaII/Fe]} & $-$0.59 $\pm$ 0.61 (2) &    --  &    --  &    --  &    --  \\
{[CaI/Fe]} & $-$0.02 $\pm$ 0.08 (7) &  0.04 $\pm$ 0.06 (3) & $-$0.05 $\pm$ 0.08 (5) &  0.00 $\pm$ 0.07 (15) &  0.06 $\pm$ 0.19 (4) \\
{[CeII/Fe]} & $-$0.21 $\pm$ 0.06 (2) &  0.10 $\pm$ 0.13 (5) & $-$0.01 $\pm$ 0.14 (3) & $-$0.06 $\pm$ 0.03 (3) &  0.02 $\pm$ 0.13 (4) \\
{[CoI/Fe]} &  0.13 $\pm$ 0.12 (12) &  0.05 $\pm$ 0.12 (16) &  0.00 $\pm$ 0.08 (10) & $-$0.09 $\pm$ 0.11 (15) &  0.02 $\pm$ 0.06 (8) \\
{[CrI/Fe]} & $-$0.02 $\pm$ 0.08 (19) &  0.03 $\pm$ 0.12 (16) & $-$0.04 $\pm$ 0.11 (11) & $-$0.05 $\pm$ 0.07 (13) &  0.08 $\pm$ 0.10 (10) \\
{[CrII/Fe]} & $-$0.09 $\pm$ 0.00 (1) & $-$0.03 $\pm$ 0.00 (1) &    --  & $-$0.13 $\pm$ 0.00 (1) & $-$0.10 $\pm$ 0.00 (1) \\
{[CuI/Fe]} & $-$0.13 $\pm$ 0.03 (2) &  0.05 $\pm$ 0.45 (2) & $-$0.36 $\pm$ 0.04 (2) & $-$0.18 $\pm$ 0.25 (2) & $-$0.08 $\pm$ 0.00 (1) \\
{[EuII/Fe]} &  0.02 $\pm$ 0.23 (2) &  0.03 $\pm$ 0.06 (2) & $-$0.01 $\pm$ 0.09 (2) & $-$0.04 $\pm$ 0.21 (2) & $-$0.06 $\pm$ 0.01 (2) \\
{[LaII/Fe]} & $-$0.35 $\pm$ 0.03 (4) &  0.03 $\pm$ 0.03 (2) & $-$0.12 $\pm$ 0.00 (2) & $-$0.21 $\pm$ 0.08 (4) & $-$0.19 $\pm$ 0.06 (4) \\
{[MgI/Fe]} &  0.14 $\pm$ 0.05 (2) &  0.15 $\pm$ 0.01 (2) &  0.23 $\pm$ 0.06 (2) &  0.13 $\pm$ 0.05 (2) &  0.27 $\pm$ 0.01 (2) \\
{[NaI/Fe]} &  0.08 $\pm$ 0.05 (4) &  0.23 $\pm$ 0.03 (2) &  0.04 $\pm$ 0.38 (2) &  0.14 $\pm$ 0.05 (3) &  0.48 $\pm$ 0.00 (1) \\
{[NiI/Fe]} &  0.02 $\pm$ 0.12 (23) & $-$0.03 $\pm$ 0.07 (11) & $-$0.06 $\pm$ 0.08 (18) & $-$0.13 $\pm$ 0.06 (20) &  0.00 $\pm$ 0.11 (11) \\
{[ScI/Fe]} & $-$0.06 $\pm$ 0.01 (3) & $-$0.11 $\pm$ 0.12 (5) & $-$0.10 $\pm$ 0.05 (3) & $-$0.13 $\pm$ 0.01 (3) & $-$0.10 $\pm$ 0.08 (3) \\
{[ScII/Fe]} &  0.02 $\pm$ 0.08 (7) &  0.10 $\pm$ 0.12 (4) &  0.10 $\pm$ 0.13 (4) & $-$0.01 $\pm$ 0.08 (8) &  0.03 $\pm$ 0.10 (3) \\
{[SI/Fe]} &    --  &    --  &  0.83 $\pm$ 0.00 (1) &    --  &    --  \\
{[SiI/Fe]} & $-$0.03 $\pm$ 0.08 (12) &  0.07 $\pm$ 0.10 (10) &  0.02 $\pm$ 0.07 (9) & $-$0.08 $\pm$ 0.08 (11) &  0.17 $\pm$ 0.14 (7) \\
{[SiII/Fe]} &    --  &  0.41 $\pm$ 0.00 (1) &    --  & $-$0.31 $\pm$ 0.10 (2) &  0.93 $\pm$ 0.00 (1) \\
{[SrI/Fe]} &  0.26 $\pm$ 0.00 (1) &  0.73 $\pm$ 0.26 (2) &  0.79 $\pm$ 0.00 (1) &    --  &  0.77 $\pm$ 0.17 (2) \\
{[TiI/Fe]} &  0.09 $\pm$ 0.07 (45) &  0.04 $\pm$ 0.07 (35) &  0.02 $\pm$ 0.08 (34) &  0.02 $\pm$ 0.08 (50) &  0.14 $\pm$ 0.12 (21) \\
{[TiII/Fe]} &  0.07 $\pm$ 0.12 (11) &  0.04 $\pm$ 0.14 (7) &  0.02 $\pm$ 0.10 (4) &  0.01 $\pm$ 0.10 (9) &  0.04 $\pm$ 0.20 (6) \\
{[VI/Fe]} &  0.12 $\pm$ 0.11 (18) &  0.10 $\pm$ 0.06 (12) & $-$0.02 $\pm$ 0.10 (17) & $-$0.05 $\pm$ 0.07 (18) &  0.16 $\pm$ 0.02 (5) \\
{[YII/Fe]} & $-$0.23 $\pm$ 0.08 (7) & $-$0.07 $\pm$ 0.04 (6) & $-$0.15 $\pm$ 0.11 (7) & $-$0.12 $\pm$ 0.16 (10) & $-$0.09 $\pm$ 0.17 (6) \\
{[ZnI/Fe]} &  0.03 $\pm$ 0.54 (2) & $-$0.42 $\pm$ 0.00 (1) &  0.08 $\pm$ 0.65 (2) & $-$0.16 $\pm$ 0.29 (2) &  1.32 $\pm$ 0.00 (1) \\
{[ZrI/Fe]} &  0.01 $\pm$ 0.06 (5) &  0.17 $\pm$ 0.02 (6) &  0.12 $\pm$ 0.02 (2) &    --  &  0.06 $\pm$ 0.12 (6) \\
{[ZrII/Fe]} &  0.26 $\pm$ 0.12 (2) &  0.52 $\pm$ 0.11 (2) &    --  &  0.32 $\pm$ 0.11 (2) &  0.56 $\pm$ 0.13 (2) \\
\hline
\end{tabular}
\end{sidewaystable*}

\end{appendix}

\end{document}